\documentclass[a4paper,11pt]{article}
\pdfoutput=1 

\usepackage{jcappub} 

\usepackage[T1]{fontenc} 
\usepackage{graphicx}
\usepackage{subfigure}
\usepackage{txfonts}
\usepackage{amsmath}
\usepackage{amssymb}
\usepackage{mathtools}
\usepackage{braket}
\usepackage{esint}
\usepackage{bm}

\title{Constraints on self-interacting Bose-Einstein condensate dark matter using large-scale observables}

\author[1]{S. T. H. Hartman\note{Corresponding author.},}
\author{H. A. Winther}
\author{and D. F. Mota}

\affiliation{Institute of Theoretical Astrophysics, University of Oslo, PO Box 1029, Blindern 0315, Oslo, Norway}

\emailAdd{s.t.h.hartman@astro.uio.no}
\emailAdd{h.a.winther@astro.uio.no}
\emailAdd{d.f.mota@astro.uio.no}

\abstract{Constraints on the cosmic history of self-interacting Bose-Einstein condensed (SIBEC) dark matter (DM) are obtained using the cosmic microwave background (CMB), baryonic acoustic oscillations (BAO), growth factor measurements, and type Ia supernovae (SNIa) distances. Four scenarios are considered, one with purely SIBEC-DM, and three in which SIBEC-DM is the final product of some transition from different initial states, which are either cold, warm, or has a constant equation of state. Using a fluid approximation for the self-interacting scalar field it is found that in the simplest scenario of purely SIBEC-DM the self-interaction necessary for solving the cusp-core problem, with core-radii of low-mass halos of order $R_c\gtrsim 1\text{kpc}$, is excluded at $2.4\sigma$, or $98.5\%$ confidence. Introducing a transition, however, relaxes this constraint, but the transitions are preferred to be after matter-radiation equality, and the initial phase to be cold.}


\begin{document}

\maketitle
\flushbottom
   
\section{Introduction}
In the cosmological concordance model---the $\Lambda$CDM model---the majority of matter in the universe is in the form of cold DM (CDM), which only interacts with itself and other particles through gravity. The CDM phenomenology provides excellent agreement with many observables, ranging from the dynamics of galaxies, the formation of large scale structure, the cosmic microwave background (CMB), and the expansion history of the universe \cite{Davis1985,Percival2001,Tegmark2004,Trujillo-Gomez2011,Vogelsberger2014,Planck2015,Riess2016}. However, cosmological $N$-body simulations, assuming CDM, predict a high abundance of low-mass halos, in conflict with the small number of observed satellites in the Local Group \cite{Garrison-Kimmel2014}. The Milky Way also seems to be missing massive subhaloes that should be too big to fail at forming bright satellites, which suggest that there are either unknown or poorly understood processes that keep these massive subhaloes dark, or that suppress their formation and survival \cite{Boylan-Kolchin2011}. Furthermore, CDM halos in simulations are well-fitted by the NFW profile, which has a central cusp in the density profile, going like $\rho(r) \sim r^{-1}$ near the center, while observations of low-surface brightness galaxies and dwarf galaxies favor instead a flatter central profile (see e.g. refs. \cite{Weinberg2015,DelPopolo2017,Bullock2017}, and references therein). These issues, often referred to as the "missing satellite", "too-big-to-fail", and "cusp-core" problems, respectively, may simply be due to an incomplete knowledge of how baryonic effects should be implemented in numerical simulations to faithfully capture the underlying physics. The practical impossibility of including all physical scales needed for understanding the formation and evolution of small-scale structures makes it necessary to use sub-grid models for baryonic processes, such as star formation and supernovae feedback. It may also be that current observations have not yet found and identified the low-mass halos and ultra-faint galaxies needed to resolve some of these issues, due to their small size and low luminosity.

Another possibility is that DM is responsible for these small-scale discrepancies, as most properties of DM are completely unknown, and its particle identity yet to be determined. There is no shortage of DM candidates in the literature, popular examples being weakly interacting massive particles (WIMPs), ultra-light scalars, primordial black holes, and sterile neutrinos \cite{Roszkowski2018, Boyarsky2019, Ferreira2020, Villanueva-Domingo2021}. Some of these are largely motivated by extensions to the Standard Model (SM) of particle physics. WIMPs, for instance, arise naturally in many beyond the SM scenarios, but also give the correct relic DM density if they were thermally produced, with coupling strengths comparable to the weak interaction. Direct detections searches, however, have yet to find a reliable WIMP signal, and are approaching the neutrino floor where the neutrino background will impede further progress in constraining the WIMP parameter space \cite{Monroe2007, Billard2014, Ohare2020}. Sterile neutrinos are another type of hypothetical particles that might explain the non-zero mass of SM neutrinos, but only interact with SM particles through gravity, and might therefore contribute to DM \cite{Boyarsky2019}. Ultra-light scalar, or pseudo-scalar, particles appear frequently in particle physics theories beyond the SM, such as the original axion, proposed as a solution to the strong-$CP$ problem in quantum chromodynamics (QCD), or in string theory where axion-like particles are a generic prediction \cite{Marsh2016}. If the mass of such a light scalar DM particle is sufficiently small, its de Broglie wavelength $\lambda_{\text{dB}}=1/mv$ can reach astrophysical size, and small-scale structure is suppressed due to its wave-like behavior across galactic distances, though the small mass also means that these particles would have to have been produced non-thermally \cite{Ferreira2020}. This type of DM is more generally termed Fuzzy DM (FDM), because the particles can be regarded as "fuzzy" due to the Heisenberg uncertainty principle. Cosmological simulations have shown that FDM halos have solitonic cores, and approaches the NFW profile at larger radii \cite{Schive2014,Mocz2017,Nori2020,Mina2020b,May2021}. Simulations and cosmological data suggest the mass of the ultra-light scalar particle in the FDM scenario to be $10^{-22}$ -- $10^{-21}\text{eV}$ \cite{Hlozek2015,Irsic2017,Kobayashi2017,Armengaud2017,Hlozek2018,Lague2021}, around what is needed for possibly solving the small-scale problems of $\Lambda$CDM, though a recent bound from the Lyman-alpha forest places the allowed mass of the FDM particle at $m>2\times 10^{-20}\text{eV}$, disfavoring the canonical FDM mass range \cite{Rogers2021}.

A large de Broglie wavelength is not the only way for ultra-light DM to flatten the central regions of low-mass halos. Because of their small particle mass, $m<1\text{eV}$, and therefore very high number densities, ultra-light DM need only have a small self-interaction to have a Jeans' length on the order of $\text{kpc}$, such that halo cores are supported by an interaction pressure \cite{Peebles2000,Goodman2000,Arbey2002,Arbey2003,Bohmer2007,Chavanis2011,Rindler-Daller2012}. We will refer to this kind of ultra-light scalar DM as self-interacting Bose-Einstein condensed DM, or SIBEC-DM, since a common feature in these models is that the DM component can be regarded as being in a Bose-Einstein condensed (BEC) phase, in which a large fraction of the bosonic DM particles are in the ground state, exhibiting long-range coherence, and can be described by a single classical wavefunction. They are also referred to as superfluid DM in the literature, due to the close relationship between SIBEC systems and superfluidity \cite{Pitaevskii2016}. Semi-analytical calculations have shown that slowly rotating hydrostatic SIBEC-DM halos can reproduce rotation curves in dwarf galaxies \cite{Zhang2018,Craciun2020}, and that the resulting halo core radii are independent of the core density, given only by the particle mass and self-interaction \cite{Chavanis2011}. There are also more exotic SIBEC-DM scenarios in the literature, in which, for example, MONDian fifth forces between baryons inside galaxies are mediated by excitations in the SIBEC-DM fluid \cite{Berezhiani2015,Khoury2016}, or where dark energy is an emergent phenomenon of the interaction between multiple scalar DM components \cite{Ferreira2019}.

While there have been several studies of FDM using large-scale cosmological simulations and constraints obtained from cosmological observables, similar work on SIBEC-DM has yet to be carried out. Some bounds on the SIBEC-DM parameters have been found using the background evolution, e.g. by demanding that Big Bang Nucleosynthesis (BBN) and matter-radiation equality not be drastically different from $\Lambda$CDM, which find that SIBEC-DM core radii of $R_c\approx 1\text{kpc}$ is not ruled out \cite{Arbey2002,Arbey2003,Li2014}. However, stronger bounds are expected if we go beyond the background evolution of SIBEC-DM, and consider linear perturbations, from which observables such as the CMB can be computed. This may also help constrain scenarios in which SIBEC-DM is assumed to be the end product of some transition from an earlier phase that leaves the background history largely unchanged from $\Lambda$CDM, especially if the earlier phase is very cold. Variants of such scenarios have been considered in several studies \cite{Harko2011,Harko2012,Velteen2012,Freitas2013,Bettoni2014,Freitas2015}, and it is usually assumed that DM is initially a normal gas, with a kinetic gas pressure due to a non-zero velocity dispersion, and that once the pressure of the BEC phase becomes smaller than that of the normal phase, it undergoes a phase transitions, though the mechanism that drives such a transition remains unclear. Furthermore, the DM velocity dispersion of the initial normal phase is often taken to be very small and constant, so that it is essentially cold, contrary to the expectation that particle momenta decreases with expansion as they are redshifted, with the pressure changing accordingly. A consequence of this is that DM should be warm (WDM), even hot and radiation-like, at sufficiently early times, which would leave imprints in e.g. the CMB.

In this paper we consider four scenarios; (i) pure SIBEC-DM that has an early radiation-like epoch before becoming non-relativistic, (ii) SIBEC-DM that is initially in a phase with a constant equation of state, (iii) SIBEC-DM that is initially in a warm phase, and (iv) SIBEC-DM that is initially CDM-like. The first scenario follows naturally from a scalar field Lagrangian with quartic self-interactions \cite{Matos2001,Li2014}, while the others are phenomenological models for DM that transitions from an initially cold or warm DM fluid to SIBEC-DM. Scenario (ii) in particular has been considered frequently \cite{Harko2011,Harko2012,Velteen2012,Freitas2013,Freitas2015}, while (iii) is an extension of (ii) that includes the redshifting of particle momenta in the initial phase. Two approaches are employed to model these transitions; a single-fluid approach, in which the DM equation of state and sound speed goes from the initial phase to the SIBEC phase in a smooth fashion; and a two-fluid approach, where DM starts out in the initial phase, but is converted into a separate SIBEC-DM component that coexists with the initial DM fluid. The linear equations for these scenarios are implemented into the Boltzmann code CLASS \cite{Blas2011}, and the Markov Chain Monte Carlo (MCMC) code MontePython \cite{Audren2012,Brinckmann2018} is used to place constraints on the model parameters, using CMB temperature and polarization power spectra from the Planck 2018 data release \cite{Planck2020}, BAO and growth rate measurements from BOSS \cite{Alam2017}, and the Pantheon compilation of SNIa distances \cite{Scolnic2018}. 

The paper is structured as follows: some of the basic features of ultra-light DM, with an emphasis on the SIBEC-DM models relevant for this work, are introduced in section \ref{sec:basics_SIBEC_DM}. Equations for the evolution of SIBEC-DM at the linear and background level, derived from cosmological perturbation theory, are given in section \ref{sec:cosmological_pert_theory}, with its effects on the matter power spectrum and the CMB discussed in section \ref{sec:effect_of_SIBEC_DM_on_observables}. The main results of this paper, the constraints on the phenomenology of the SIBEC-DM models obtained from large-scale observables, are presented in section \ref{sec:constrains_from_cosmological_data}, and final conclusions given in section \ref{sec:conclusions}.

\section{Self-interacting Bose-Einstein condensates as dark matter}
\label{sec:basics_SIBEC_DM}

Ultra-light scalar DM can be described by a classical field theory, for instance by the Lagrangian for a complex scalar field $\Psi$ with self-interactions \cite{Li2014},
\begin{equation}
\label{eq:complex_interacting_klein_gordon}
    \mathcal{L} = \frac{1}{2m}g_{\mu\nu}\partial^{\mu}\Psi^{*}\partial^{\nu}\Psi - \frac{1}{2}m|\Psi|^2 - \frac{1}{2}g|\Psi|^4.
\end{equation}
In the late universe, where the interaction energy density of the scalar field is small compared to its rest mass, the non-relativistic limit can be used to accurately describe the field, and is obtained by defining the non-relativistic field $\psi$,
\begin{equation}
    \Psi = \psi e^{-imt},
\end{equation}
and neglecting terms that contain oscillating factors of $e^{-imt}$ \cite{Ferreira2020};
\begin{equation}
\label{eq:gross_pitaevskii}
    i\frac{\partial \psi}{\partial t} = \Bigg[\frac{-\nabla^2}{2m} + g|\psi|^2 + mV_{\text{ext}}\Bigg]\psi.
\end{equation}
This equation is called the Gross-Pitaevskii equation, or the non-linear Schrödinger equation (NLSE), and can also be derived as the zero-temperature limit of the quantum Hamiltonian of the many-body bosonic system using mean-field theory \cite{Pitaevskii2016}. The wavefunction $\psi$ is related to the particle number density by $|\psi|^2=n$, and evolves under the influence of effective two-body contact interactions, parameterized by $g$, and the external potential $V_{\text{ext}}$, taken to be the gravitational potential $\phi$. For FDM it is the kinetic term that gives rise to their "fuzzy" properties and suppresses small-scale structure. In SIBEC-DM, it is instead the interaction term, and the resulting interaction pressure, that counteracts gravity. This can be seen more clearly if the NLSE is reformulated in a hydrodynamical form by inserting for the wavefunction
\begin{equation}
\label{eq:madelung_transformation}
    \psi = \sqrt{n} e^{iS} = \sqrt{\frac{\rho}{m}} e^{iS},
\end{equation}
and defining the velocity field $\bm{v} = \bm{\nabla}S/m$. The complex NLSE can then be separated into a continuity equation for mass density, and a quantum variant of the momentum equation;
\begin{equation}
\label{eq:madelung_continuity}
    \frac{\partial \rho}{\partial t} + \bm{\nabla}\cdot(\rho \bm{v}) = 0,
\end{equation}
\begin{equation}
\label{eq:madelung_momentum}
    \frac{\partial \bm{v}}{\partial t} + (\bm{v}\cdot \bm{\nabla}) \bm{v} + \bm{\nabla}\Bigg(\frac{g\rho}{m^2} - \frac{1}{2m^2}\frac{\nabla^2 \sqrt{\rho}}{\sqrt{\rho}} + \phi \Bigg) = \bm{0}.
\end{equation}
These are called the Madelung equations \cite{Madelung1926}, and describe a dissipationless fluid with gravity, a fluid pressure due to self-interactions,
\begin{equation}
\label{eq:madelung_pressure}
    P = \frac{g\rho^2}{2m^2},
\end{equation}
and a peculiar potential due to the kinetic term in the NLSE,
\begin{equation}
\label{eq:madelung_quantum_potential}
    Q = -\frac{1}{2m^2}\frac{\nabla^2 \sqrt{\rho}}{\sqrt{\rho}}.
\end{equation}
This quantity is often called the quantum potential, and provides an effective pressure even in the absence of self-interactions. SIBEC-DM will in principle always have a contribution from the quantum potential, but for sufficiently large particle mass $m$ and interaction strength $g$, and small curvature in the density field, it is negligible compared to $P$.

Both eq. \eqref{eq:gross_pitaevskii}, and eqs. \eqref{eq:madelung_continuity} and \eqref{eq:madelung_momentum}, have been used in numerous studies of FDM and SIBEC-DM, such as cosmological simulations of FDM \cite{Schive2014,Mocz2017,Nori2018,Nori2020,Mina2020,Mina2020b,May2021}, and for investigating the properties of ultra-light DM halos \cite{Chavanis2011,Chavanis2011b,Rindler-Daller2014,Zhang2018,Craciun2020,Lancaster2020}. For SIBEC-DM, a simple, but particularly useful, constraint on the combination $g/m^2$ can be found by assuming SIBEC-DM halos to be spherical and in hydrostatic equilibrium. The resulting halo profile is \cite{Chavanis2011}
\begin{equation}
    \rho(r) = \rho_0\frac{\sin(Ar)}{Ar},
\end{equation}
where $A = \sqrt{4\pi Gm^2/g}$, and goes to zero at
\begin{equation}
\label{eq:SIBEC_DM_Rc}
    r = R_c = \sqrt{\frac{g\pi}{4Gm^2}}.
\end{equation}
An estimate of the SIBEC-DM halo core size is therefore $R_c$, and yields the correspondence
\begin{equation}
\label{eq:gm2_limit_hydrostatic_equilibrium}
    \frac{g}{m^2} = \frac{4GR_c^2}{\pi} \approx 2\times 10^{-4} \left(\frac{R_c}{1\text{kpc}}\right)^2 \text{eV}^{-4},
\end{equation}
or alternatively, if we insert the DM background density today, gives for the SIBEC-DM equation of state
\begin{equation}
\label{eq:w0_limit_hydrostatic_equilibrium}
    \omega_0 = \frac{\bar{P}_0}{\bar{\rho}_0} = \frac{g\bar{\rho}_0}{2m^2} \approx 10^{-15} \left(\frac{R_c}{1\text{kpc}}\right)^2.
\end{equation}
By fitting the rotation curves of slowly rotating SIBEC-DM halos at hydrostatic equilibrium to nearby galaxies from the Spitzer Photometry \& Accurate Rotation Curves (SPARC) database \cite{Lelli2016}, the combination $g/m^2$ is estimated to be around $3\times 10^{-4} \,\text{eV}^{-4}$ to $5\times 10^{-2} \,\text{eV}^{-4}$ \cite{Zhang2018,Craciun2020}, about the same as obtained from eq. \eqref{eq:gm2_limit_hydrostatic_equilibrium} for $R_c \approx 1 \,\text{kpc}$ to $10 \,\text{kpc}$. The SIBEC-DM self-interaction should, of course, also satisfy constraints from colliding galaxy clusters. By comparing the spatial offset of stars, gas, and DM in these collisions, the cross section $\sigma$ of DM can be bounded \citep{Harvey2015}, and the lack of deceleration of DM and its proximity to the collisionless stars places an upper limit of $\sigma/m \lesssim 0.5 \text{cm}^2/\text{g}$. This translates to a bound on the self-interaction \cite{Pitaevskii2016}
\begin{equation}
\label{eq:cluster_constraint}
    g = \frac{\sqrt{4\pi\sigma}}{m} \lesssim 5\times 10^{-12}\left(\frac{1\,\text{eV}}{m}\right)^{1/2} \,\text{eV}^{-2},
\end{equation}
which in terms of $R_c$ reads
\begin{equation}
    R_c \lesssim 2\times 10^{-4} \left( \frac{1\text{eV}}{m} \right)^{5/4} \text{kpc}.
\end{equation}

\begin{figure}
    \centering
    \includegraphics[width=0.7\linewidth]{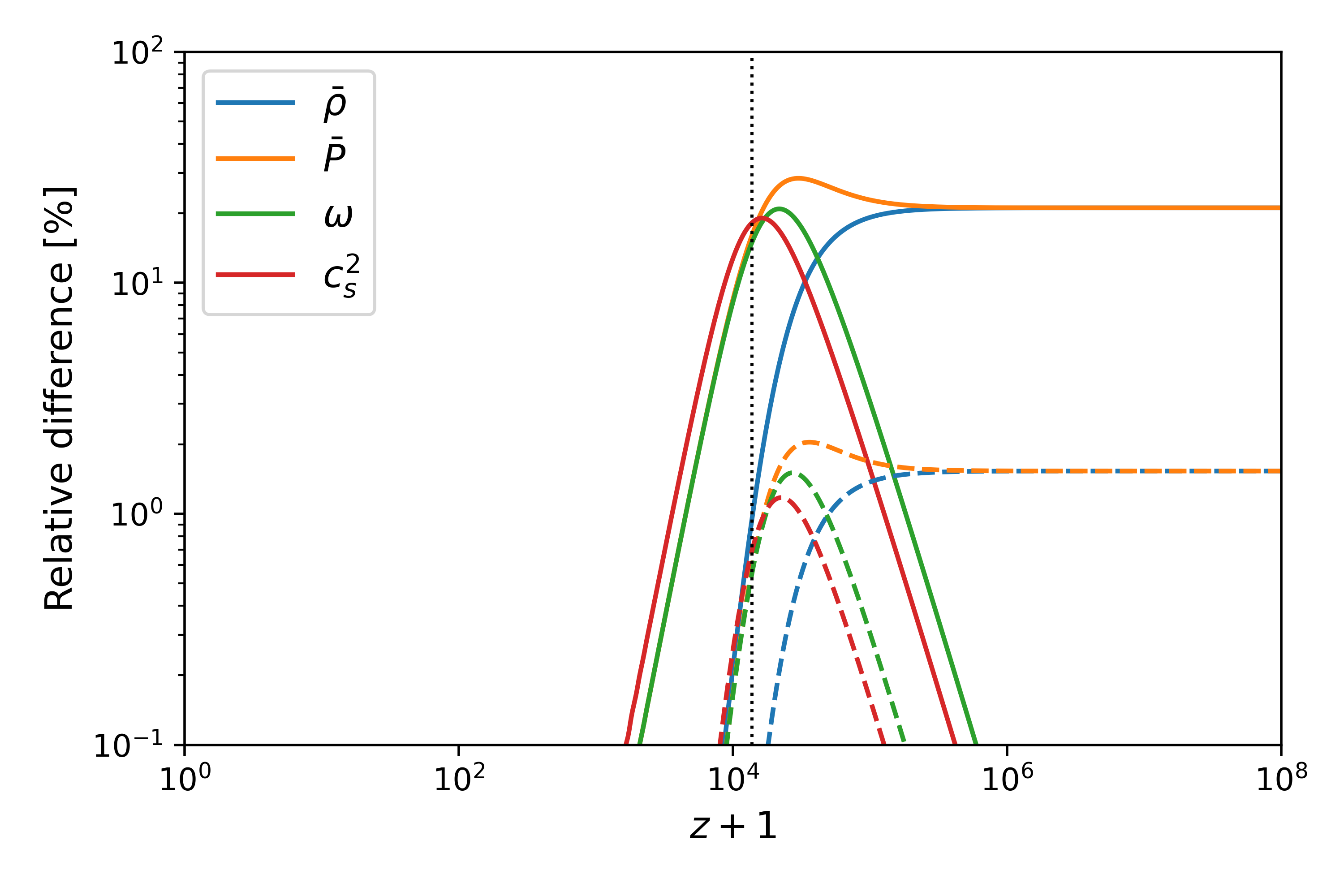}
    \caption{The relative difference of SIBEC-DM background quantities using the equation of state given by eq. \eqref{eq:SIBEC_eos} (\textit{solid lines}) and by eq. \eqref{eq:SIBEC_eos_improved} (\textit{dashed lines}) compared to eq.  \eqref{eq:pressure_fast_osc_approximation}, for $g/m^2=10^{-2}\,\text{eV}$. The vertical dotted line indicates where $\omega_{\text{SIBEC}}=0.1$, around where SIBEC-DM ceases to be radiation-like. Varying $g/m^2$ or $\bar{\rho}_0$ only changes the redshift of the crossover between radiation-like and the non-relativistic equation of state, not the magnitude of the relative difference.}
    \label{fig:eos_comparison}
\end{figure}

The NLSE and Madelung equations for SIBEC-DM work well for sufficiently low densities, but in the early universe, where the density is very high and the interaction energy dominates the total energy density of the DM scalar field, the non-relativistic approximation breaks down. To remedy this we must therefore return to the relativistic field Lagrangian. Following ref. \cite{Li2014}, directly solving eq. \eqref{eq:complex_interacting_klein_gordon} at the background level, assuming the oscillation of the field $\Psi$ to be faster than the rate of expansion, gives the averaged energy density $\bar{\rho}$ and pressure $\bar{P}$ as
\begin{equation}
    \braket{\bar{\rho}} = m\braket{|\Psi|^2} + \frac{3}{2}g\braket{|\Psi|^2}^2,
\end{equation}
\begin{equation}
    \braket{\bar{P}} = \frac{1}{2}g\braket{|\Psi|^2}^2,
\end{equation}
such that
\begin{equation}
    \label{eq:pressure_fast_osc_approximation}
    \braket{\bar{P}} = \frac{m^2}{18g}\left(\sqrt{1 + \frac{6g\braket{\bar{\rho}}}{m^2}} - 1 \right)^2.
\end{equation}
This expression reduces to the non-relativistic pressure in eq. \eqref{eq:madelung_pressure} in the limit $\braket{\bar{\rho}}\ll m^2/g$, upon equating the energy density to mass density. The limit $\braket{\bar{\rho}}\gg m^2/g$, where the self-interaction dominates the SIBEC-DM energy density, gives instead
\begin{equation}
    \braket{\bar{P}} = \frac{1}{3}\braket{\bar{\rho}},
\end{equation}
hence the complex scalar field DM behaves like radiation at high densities. Ref. \citep{Li2014} identified also a very early stiff phase with $P=\rho$, when the expansion rate is larger than field oscillations, though in this work we will neglect this initial epoch and only focus on the radiation-like and non-relativistic phases. These phases are also present in real scalar fields with self-interactions \cite{Matos2001}. As a  simplification, an equation of state for SIBEC-DM that is only dependent on the scale factor is used in this work,
\begin{equation}
    \label{eq:SIBEC_eos}
    \omega_{\text{SIBEC}} = \frac{1}{3}\frac{1}{1 + a^3/3\omega_0},
\end{equation}
where $\omega_0 = g\bar{\rho}_0/2m^2$ is the non-relativistic equation of state today. The resulting energy density is
\begin{equation}
 \label{eq:SIBEC_rho}
    \bar{\rho} = \bar{\rho}_0 a^{-4}\Bigg( \frac{3\omega_0 + a^3}{3\omega_0 + 1} \Bigg)^{1/3}.
\end{equation}
The same is done for the sound speed,
\begin{equation}
    \label{eq:SIBEC_cs2}
    c_{s,\text{SIBEC}}^2 = \frac{\partial \bar{P}}{\partial \bar{\rho}} = \frac{1}{3}\frac{1}{1 + a^3/3c_{s0}^2},
\end{equation}
where $c_{s0}^2 = 2\omega_0$. These simplified expressions for SIBEC-DM deviate around $20\%$-$30\%$ from eq. \eqref{eq:pressure_fast_osc_approximation} in the radiation-like era and the crossover to the non-relativistic regime, as illustrated in figure~\ref{fig:eos_comparison}. A better approximation, accurate to within $2\%$-$3\%$, can be obtained by inserting eq. \eqref{eq:SIBEC_rho} back into the pressure, eq. \eqref{eq:pressure_fast_osc_approximation}, which gives
\begin{equation}
\label{eq:SIBEC_eos_improved}
    \omega_{\text{SIBEC}} = \frac{1}{3R}(\sqrt{1 + R}-1)^2,
\end{equation}
\begin{equation}
\label{eq:SIBEC_cs2_improved}
    c_{s,\text{SIBEC}} = \frac{1}{3}(1 - (1 + R)^{-1/2}),
\end{equation}
where
\begin{equation}
    R = 12\omega_0 a^{-4}\Bigg( \frac{3\omega_0 + a^3}{3\omega_0 + 1} \Bigg)^{1/3}.
\end{equation}
However, this does not yield a simple expression for $\rho$, and the improved treatment only results in a minor change in the large-scale observables, with negligible impact on the constraints on $\omega_0$, as we shall see in section \ref{sec:constrains_from_cosmological_data}. We therefore only use the better approximation eq. \eqref{eq:SIBEC_eos_improved} when placing constraints on purely SIBEC-DM, and use the simpler equation of state \eqref{eq:SIBEC_eos} when transitions are included.

It has been suggested that SIBEC-DM has not necessarily always been in the state described by a classical field, but has undergone a transition at some point \cite{Silverman2002,Harko2011,Harko2012,Velteen2012,Freitas2013,Bettoni2014,Freitas2015}, though the mechanism for such a transition in several of these scenarios is unclear, especially since a cosmological boson gas that is below its critical temperature today is expected to have been so on the past as well, and therefore to have entered the BEC phase in the very early universe \cite{Sikivie2009,Erken2012,Saikawa2013,Guth2015}. In this work we do not seek to propose a concrete mechanism, but instead model the transition phenomenologically in two different ways; as a single fluid, in which the DM fluid has an effective equation of state that interpolates from the initial phase to the late-time SIBEC-DM fluid at the time of the transition; and as two separate, coexisting DM components, with all of DM starting out in the initial phase, and is eventually converted into a SIBEC-DM fluid. In the following few sections we describe the scenarios considered in this work, with an emphasis on the single-fluid approach.

\subsection{SIBEC-DM transitioning from a fluid with constant equation of state}
\label{sec:sigma02_SIBEC_DM}

Refs. \cite{Harko2011,Harko2012,Velteen2012,Freitas2013,Freitas2015} suggest that SIBEC-DM was initially in an uncondensed phase---a normal fluid---with a gas pressure given by
\begin{equation}
    P = \frac{1}{3}\braket{v^2}\rho,
\end{equation}
where $\braket{v^2}$ is the average squared velocity of DM, taken to be constant. The pressure and energy density is hence
\begin{equation}
    P = \sigma_0^2\,\rho,
\end{equation}
and
\begin{equation}
\label{eq:rho_constant_w_SIBEC}
    \bar{\rho} = \bar{\rho}_0 a^{-3(1+\sigma_0^2)},
\end{equation}
with $\sigma_0^2 = \braket{v_0^2}/3 \ll 1/3$. The transition into the condensed phase is then assumed to take place once the SIBEC interaction pressure is smaller than the normal gas pressure, i.e. the DM fluid settles into the phase with the lowest pressure and preserves continuity in the pressure. The transition point is therefore to a good approximation \cite{Harko2011}
\begin{equation}
\label{eq:constant_w_SIBEC_transition_point}
    a_{t} = (\omega_0/\sigma_0^2)^{1/3},
\end{equation}
To model the transition of the DM fluid from the initial gas-like phase to the SIBEC fluid in the single-fluid approach, we use the equation of state
\begin{equation}
\label{eq:eq:constant_w_SIBEC_eos}
    \omega_{\chi} = \Big[\Big(\sigma_0^2\Big)^{-\kappa} + \Big(\omega_0 a^{-3}\Big)^{-\kappa}\Big]^{-1/\kappa}.
\end{equation}
This essentially picks the smallest pressure, with a crossover region centered at $a_t$ and has a width determined by $\kappa$, with a large $\kappa$ corresponding to a nearly instantaneous transition. The special case $\kappa=1$ has the simple expression for the energy density
\begin{equation}
    \bar{\rho} = \bar{\rho}_0 a^{-3}\Bigg(\frac{\sigma_0^2 + \omega_0 a^{-3}}{\sigma_0^2 + \omega_0} \Bigg)^{\sigma_0^2}.
\end{equation}
The sound speed is modelled in the same way,
\begin{equation}
\label{eq:eq:constant_w_SIBEC_cs2}
    c^2_{s,\chi} = \Big[\Big(\sigma_0^2\Big)^{-\kappa} + \Big(c^2_{s0} a^{-3}\Big)^{-\kappa}\Big]^{-1/\kappa}.
\end{equation}

\subsection{SIBEC-DM transitioning from warm DM}
\label{sec:warm_SIBEC_DM}

The above scenario completely neglects the redshifting of particle momenta, and a more realistic equation of state for the initial normal fluid phase is therefore desirable. The so-called reduced relativistic gas (RRG) approximation is well suited for this purpose. RRG assumes that all the particles of a relativistic gas have equal velocities, and the resulting equation of state is accurate to within $2.5\%$ of realistic relativistic ideal gases and reproduces well the effect of warm DM on large-scale observables \cite{Sakharov1966, Berredo-Peixoto2005, Fabris2009, Fabris2012, Medeiros2012, Reis2018, Hipolito-Ricaldi2018, Silva2019}. We refer the reader to refs. \cite{Berredo-Peixoto2005, Silva2019} for details on RRG, and simply state the resulting equation of state here;

\begin{equation}
    \omega = c_s^2 = \frac{1}{3}\frac{b^2}{a^2 + b^2}.
\end{equation}
The parameter $b$ gives the warmness of the DM gas at the present time, and is roughly equal to the average DM velocity today $b^2\approx \braket{v^2_0}$ for $b\ll 1$, giving $\omega\sim a^{-2}$ in the late-time non-relativistic era, and $\omega=1/3$ in the early relativistic era, as expected. The energy density in the RRG approximation is given by a simple expression,
\begin{equation}
\label{eq:rho_warm_SIBEC}
    \bar{\rho} = \bar{\rho}_0 a^{-3}\sqrt{\frac{1 + b^2 a^{-2}}{1 + b^2}}.
\end{equation}
In the context of thermal relics, a warmness parameter of $b^2\approx 2\times 10^{-15}$ corresponds to a mass $m=3.5\text{keV}$, and $b^2\approx 7\times 10^{-15}$ to $m=5.3\text{keV}$ \cite{Hipolito-Ricaldi2018}.

We consider this scenario only in the two-fluid approach, detailed in section \ref{sec:cosmological_pert_theory}, since the single-fluid approach is very similar to the two-fluid approach with fast transitions, regardless of rate of the transition.

\subsection{SIBEC-DM transitioning from cold DM}
\label{sec:cold_SIBEC_DM}

Finally, we consider a purely phenomenological scenario in which DM transitions from a CDM-like initial state to a SIBEC-DM fluid,
\begin{equation}
\label{eq:cold_SIBEC_eos}
    \omega_{\chi} = \frac{(a/a_t)^{\kappa}}{1 + (a/a_t)^{\kappa}} \omega_{\text{SIBEC}},
\end{equation}
\begin{equation}
\label{eq:cold_SIBEC_cs2}
    c^2_{s,\chi} = \frac{(a/a_t)^{\kappa}}{1 + (a/a_t)^{\kappa}} c^2_{s,\text{SIBEC}}.
\end{equation}
The transition time $a_t$ is in this case not dependent on any parameter of the initial and final phases, as was assumed in the two previous scenarios. The resulting constraints therefore represents a rough upper bound on the self-interaction of SIBEC-DM as a function of transition time and rate, since we usually expect modifications to DM to provide a worse fit to large-scale observables compared to CDM.

\section{Cosmological perturbation theory}
\label{sec:cosmological_pert_theory}
We use cosmological perturbation theory for computing the growth of structure from primordial adiabatic initial conditions and finding how large-scale observables are affected by SIBEC-DM. In the synchronous gauge the metric tensor is given by
\begin{equation}
    \text{d}s^2 = g_{\mu\nu}\text{d}x^{\mu}\text{d}x^{\nu} = a^2[-\text{d}\tau^2 + (\delta_{ij} + h_{ij})\text{d}x^{i}\text{d}x^{j}],
\end{equation}
where $\tau$ is conformal time, and $h_{ij}$ metric perturbations \cite{Ma1995}. SIBEC-DM lives on this metric, and is in this work treated within a fluid framework \cite{Faraoni2012}, with the fluid pressure obtained by averaging over the oscillations of the background field, which gives eq. \eqref{eq:pressure_fast_osc_approximation} \cite{Li2014}. A similar approach has been used before for the free field limit, i.e. FDM \cite{Hwang2009,Park2012,Hlozek2015,Marsh2016,Hlozek2018}.
SIBEC-DM with transitions are treated as either effectively one component $\chi$ in the single-fluid scenario, or two separate components $\chi_1$ and $\chi_2$ in the two-fluid scenario. Because the latter case is more general, it is considered first.

\subsection{Two-fluid DM}
\label{sec:two_fluid_DM}
The energy-momentum tensors of two DM fluid components are \cite{Ma1995}
\begin{equation}
    T_{\chi_1}^{\mu\nu} = (\rho_{\chi_1} + P_{\chi_1})u_{\chi_1}^{\mu}u_{\chi_1}^{\nu} + P_{\chi_1}g^{\mu\nu} + \pi_{\chi_1}^{\mu\nu},
\end{equation}
\begin{equation}
    T_{\chi_2}^{\mu\nu} = (\rho_{\chi_2} + P_{\chi_2})u_{\chi_2}^{\mu}u_{\chi_2}^{\nu} + P_{\chi_2}g^{\mu\nu} + \pi_{\chi_2}^{\mu\nu},
\end{equation}
that satisfy the combined covariant conservation law
\begin{equation}
    \nabla_\nu (T_{\chi_1}^{\mu\nu} + T_{\chi_2}^{\mu\nu}) = 0,
\end{equation}
or, if a conversion function $Q$ is introduced \cite{Bringmann2018},
\begin{equation}
    \nabla_\nu T_{\chi_1}^{\mu\nu} = -\nabla_\nu T_{\chi_2}^{\mu\nu} = -Qu_{\chi_1}^{\mu}.
\end{equation}
Expanding to linear order, $\rho = \bar{\rho} + \delta \rho$, $P = \bar{P} + \delta P$, $u_{\mu} = a(-1, v_i)$, $T^{\mu\nu} = \bar{T}^{\mu\nu} + \delta T^{\mu\nu}$, and so on, gives the evolution of the two components at the background level as
\begin{equation}
\label{eq:background_eq_ch1}
    \bar{\rho}_{\chi_1}' + 3\mathcal{H}\bar{\rho}_{\chi_1}(1 + \omega_{\chi_1}) = a\bar{Q},
\end{equation}
\begin{equation}
\label{eq:background_eq_ch2}
    \bar{\rho}_{\chi_2}' + 3\mathcal{H}\bar{\rho}_{\chi_2}(1 + \omega_{\chi_2}) = -a\bar{Q},
\end{equation}
where $'=\frac{\text{d}}{\text{d}\tau}$, $\mathcal{H}=a'/a$, and $\omega=\bar{P}/\bar{\rho}$. The Fourier components of the linear perturbations $\delta = \delta\rho/\bar{\rho}$ and $\theta = \partial_i v^i$ evolve as
\begin{equation}
\begin{split}
    \delta_{\chi_1}' &+ 3\mathcal{H}(c_{s,\,\chi_1}^2-\omega_{\chi_1})\delta_{\chi_1} + (1+\omega_{\chi_1})\Big(\theta_{\chi_1} + \frac{1}{2}h'\Big) \\
    & \quad\quad = \frac{a\bar{Q}\delta_{\chi_1}}{\bar{\rho}_{\chi_1}}-\frac{a\delta Q}{\bar{\rho}_{\chi_1}},
\end{split}
\end{equation}
\begin{equation}
\begin{split}
    \theta_{\chi_1}' &+ \mathcal{H}(1-3\omega_{\chi_1})\theta_{\chi_1} + \frac{\omega_{\chi_1}'}{1+\omega_{\chi_1}}\theta_{\chi_1} - \frac{c_{s,\,\chi_1}^2 k^2}{1+\omega_{\chi_1}}\delta_{\chi_1} + k^2 \sigma_{\chi_1} \\
    & \quad\quad = \frac{a\bar{Q}\theta_{\chi_1}}{\bar{\rho}_{\chi_1}} - \frac{a\bar{Q}\theta_{\chi_1}}{\bar{\rho}_{\chi_1}(1+\omega_{\chi_1})},
\end{split}
\end{equation}
and
\begin{equation}
\begin{split}
    \delta_{\chi_2}' &+ 3\mathcal{H}(c_{s,\,\chi_2}^2-\omega_{\chi_2})\delta_{\chi_2} + (1+\omega_{\chi_2})\Big(\theta_{\chi_2} + \frac{1}{2}h'\Big) \\
    & \quad\quad = -\frac{a\bar{Q}\delta_{\chi_2}}{\bar{\rho}_{\chi_2}}+\frac{a\delta Q}{\bar{\rho}_{\chi_2}},
\end{split}
\end{equation}
\begin{equation}
\begin{split}
    \theta_{\chi_2}' &+ \mathcal{H}(1-3\omega_{\chi_2})\theta_{\chi_2} + \frac{\omega_{\chi_2}'}{1+\omega_{\chi_2}}\theta_{\chi_2} - \frac{c_{s,\,\chi_2}^2 k^2}{1+\omega_{\chi_2}}\delta_{\chi_2} + k^2 \sigma_{\chi_2} \\
    & \quad\quad = -\frac{a\bar{Q}\theta_{\chi_2}}{\bar{\rho}_{\chi_2}} + \frac{a\bar{Q}\theta_{\chi_1}}{\bar{\rho}_{\chi_2}(1+\omega_{\chi_2})}.
\end{split}
\end{equation}
The sound speed is given, as usual, by $c_s^2 = \delta \bar{P}/\delta\bar{\rho}$, the trace of the metric perturbation is $h=h^{i}_{i}$, and $\sigma$ is the anisotropic stress, due to e.g. free-streaming. In the presence of anisotropic stress, the above equations are in general not enough to close the system, and at least two more are needed for the evolution of $\sigma_{\chi_1}$ and $\sigma_{\chi_2}$. However, since the focus is on cosmological histories in which SIBEC-DM is largely cold, and is converted into a condensed (or scalar field) phase, neither of which has anisotropic stress \cite{Hu1998}, then $\sigma = 0$. In the case of an initial warm phase, the full Boltzmann hierarchy should be used for a complete treatment of WDM, but in the present work the above perfect fluid description with the RRG equation of state is employed, which captures the main features of WDM on large scales \cite{Hipolito-Ricaldi2018}.

This description of DM is closely related to the so-called generalized DM (GDM) framework, which extends CDM with a non-zero equation of state, sound speed, and viscosity, and has been studied extensively for a variety of realisations of the DM fluid properties \cite{Hu1998, Kopp2016, Thomas2016, Thomas2019, Hipolito-Ricaldi2018, Ilic2020}.

\begin{figure}
    \centering
    \includegraphics[width=0.7\linewidth]{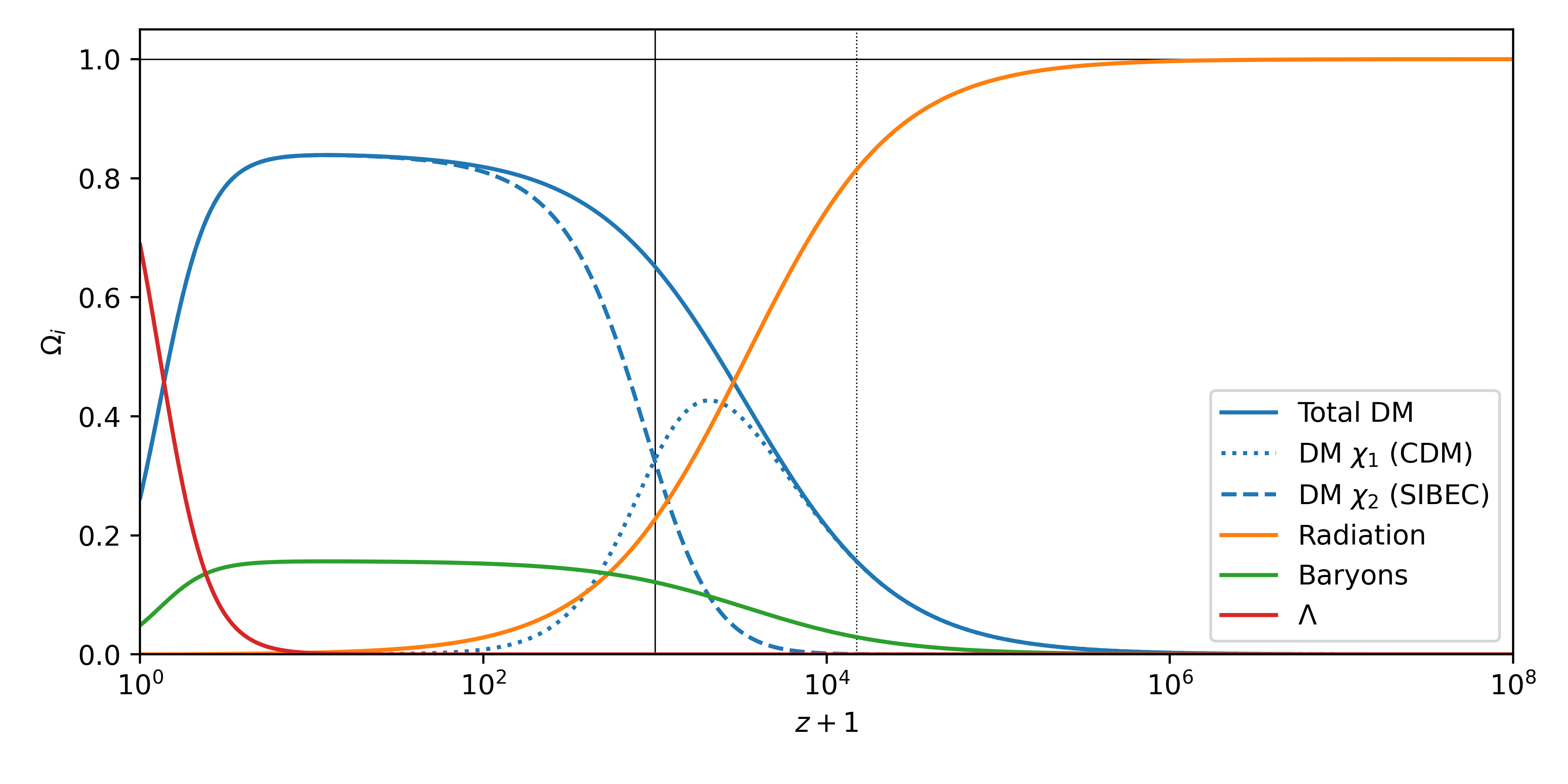}
    \includegraphics[width=0.7\linewidth]{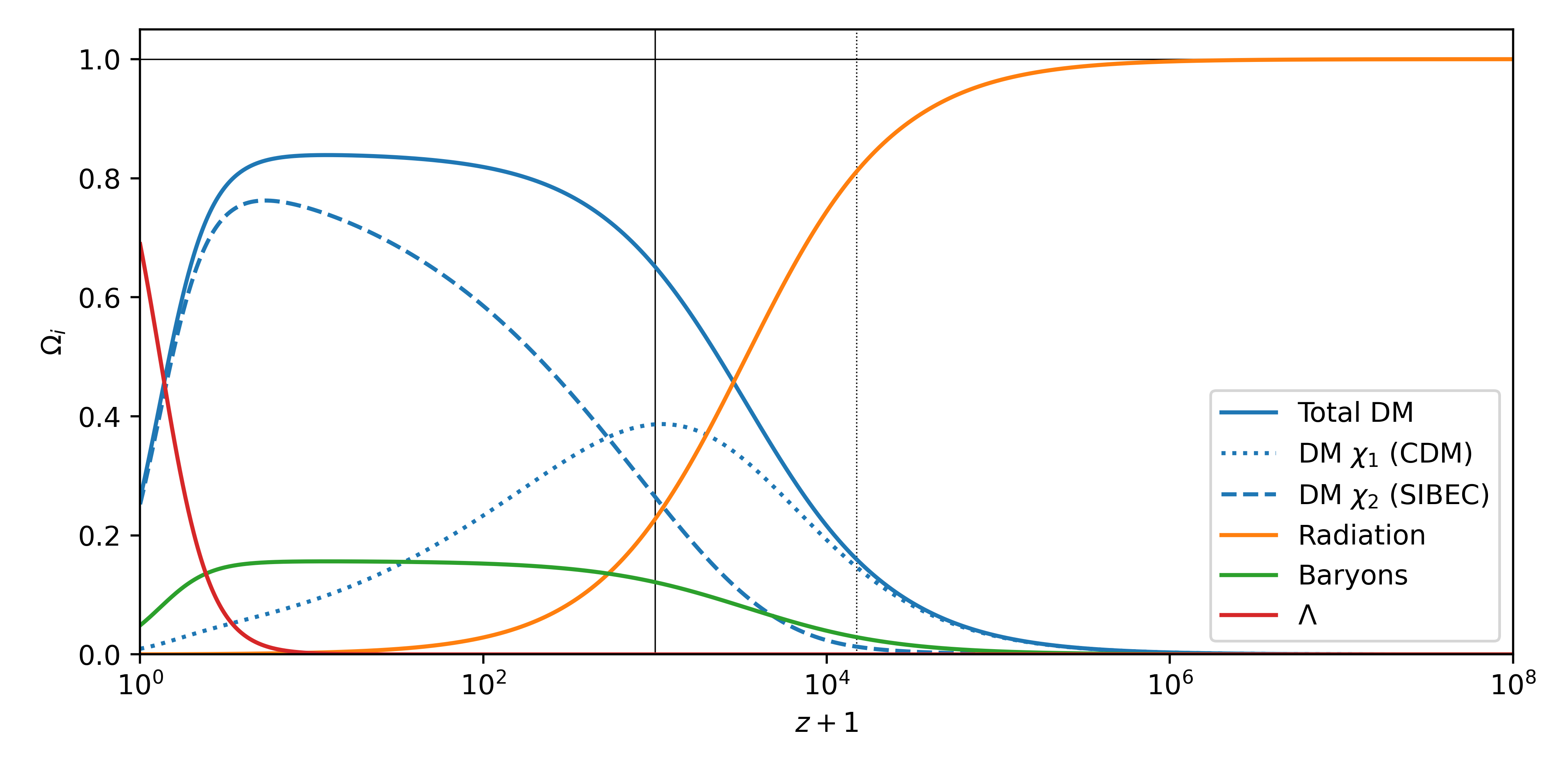}

    \caption{Relative energy densities of the universe during a cold-SIBEC transition, with $\omega_0 = 10^{-13}$, and $\zeta = 1$. The solid vertical line indicates $a_t$, while the dotted one indicates where SIBEC-DM ceases being radiation-like, $\omega_{\text{SIBEC}}=0.1$. The upper plot has $\kappa=2$ and $a_t = 10^{-3}$, and the lower has $\kappa=1/2$ and $a_t = 10^{-3}$.}
\label{fig:omegas_2fluid_examples}
\end{figure}

For the conversion factor $Q$ an approach similar to ref. \cite{Bringmann2018} is employed by assuming that $\bar{\rho}_{\chi_1}$ evolves as
\begin{equation}
\label{eq:cold_SIBEC_DM_transition_density}
    \bar{\rho}_{\chi_1} = \bar{\rho}_{i}(a) \Bigg[1 - \zeta\frac{(a/a_t)^{\kappa}}{1 + (a/a_t)^{\kappa}}\Bigg],
\end{equation}
where $\bar{\rho}_{i}$ is the background evolution of the initial phase in the absence of a transition, e.g. eqs. \eqref{eq:rho_constant_w_SIBEC} and \eqref{eq:rho_warm_SIBEC}. Written in this way, $\zeta$ gives the fraction of $\chi_1$ that is converted into $\chi_2$, $a_t$ gives the scale factor where the conversion takes place, and $\kappa$ determines the width of the transition, i.e. the rate at which component $\chi_1$ is converted into $\chi_2$. From this $\bar{Q}$ is obtained from the background equation for $\chi_1$, eq. \eqref{eq:background_eq_ch1}, and it is assumed that $\delta Q = \bar{Q}\delta_{\chi_1}$. The initial component $\chi_1$ is in the different scenarios either CDM-like with $\omega_{\chi_1}=c_{s,\chi_1}=0$, has a constant equation of state and sound speed, or is described by the RRG approximation of WDM. The second component $\chi_2$ is the SIBEC fluid, whose equation of state and sound speed are eqs. \eqref{eq:SIBEC_eos} and \eqref{eq:SIBEC_cs2}. Note that $\zeta$ generally does not correspond to the relative contribution to the cosmic energy budget of the two DM components today, because the energy density of $\chi_2$ evolves at a different rate than $\chi_1$ due to $\omega_{\chi_1} \neq \omega_{\chi_2}$, such that the resulting energy density of $\chi_2$ today is different than the fraction of $\chi_1$ that was actually converted. Throughout this work it is assumed that all of DM is eventually converted into SIBEC-DM, so $\zeta=1$. The relative contribution to the cosmic energy budget in the two-component DM approach is illustrated in figure~\ref{fig:omegas_2fluid_examples} for the cold-SIBEC transition.

\subsection{Single-fluid DM}
\label{sec:single_fluid_DM}
The effective single-fluid DM $\chi$ obeys the same equations as $\chi_1$ (and $\chi_2$, for that matter), but without any source terms due to a conversion, i.e. $Q=0$;
\begin{equation}
\label{eq:background_eq_ch}
    \bar{\rho}_{\chi}' + 3\mathcal{H}\bar{\rho}_{\chi}(1 + \omega_{\chi}) = 0,
\end{equation}
\begin{equation}
    \delta_{\chi}' + 3\mathcal{H}(c_{s,\,\chi}^2-w_{\chi})\delta_{\chi} + (1+w_{\chi})\Big(\theta_{\chi} + \frac{1}{2}h'\Big) = 0,
\end{equation}
\begin{equation}
    \theta_{\chi}' + \mathcal{H}(1-3\omega_{\chi})\theta_{\chi} + \frac{\omega_{\chi}'}{1+\omega_{\chi}}\theta_{\chi} - \frac{c_{s,\,\chi}^2 k^2}{1+\omega_{\chi}}\delta_{\chi} = 0.
\end{equation}
In this case, the two components, or phases, of DM are not treated separately, but as a tightly coupled fluid with an effective equation of state that interpolates from the value of the initial phase to the value of the final phase, as described in sections \ref{sec:sigma02_SIBEC_DM}-\ref{sec:cold_SIBEC_DM}.

We summarize the scenarios considered in this paper with the model labels that we use hereafter in table \ref{tab:model_summaries}.

\bgroup 
\def\arraystretch{1.3}
\begin{table}[t]
\caption{Brief summary of scenarios considered in this work, along with the model labels used.\newline\newline}              
\label{tab:model_summaries}      
\centering                                      
\begin{tabular}{l p{0.7\linewidth}}             
\hline\hline
Model label  & Description \\
\hline
pure SIBEC-DM       & Pure SIBEC-DM, with no transition or conversion from an earlier phase. Described in section \ref{sec:basics_SIBEC_DM}.  \\
$\sigma_0$-SIBEC-DM & DM initially in a phase with a constant equation of state, and transitions into SIBEC-DM. Considered in both the single-fluid and two-fluid approaches. Described in section \ref{sec:sigma02_SIBEC_DM}. \\
warm-SIBEC-DM       & DM initially in a warm phase described by the RRG equation of state, and transitions into SIBEC-DM. Considered only in the two-fluid approach. Described in section \ref{sec:warm_SIBEC_DM}. \\
cold-SIBEC-DM       & DM initially cold, i.e. CDM, and transitions into SIBEC-DM. Considered in both the single-fluid and two-fluid approaches. Described in section \ref{sec:cold_SIBEC_DM}.  \\
\hline\hline
\end{tabular}
\end{table}
\egroup 

\section{Effect of SIBEC-DM on observables}
\label{sec:effect_of_SIBEC_DM_on_observables}
\begin{figure}
    \centering
    \includegraphics[width=0.7\linewidth]{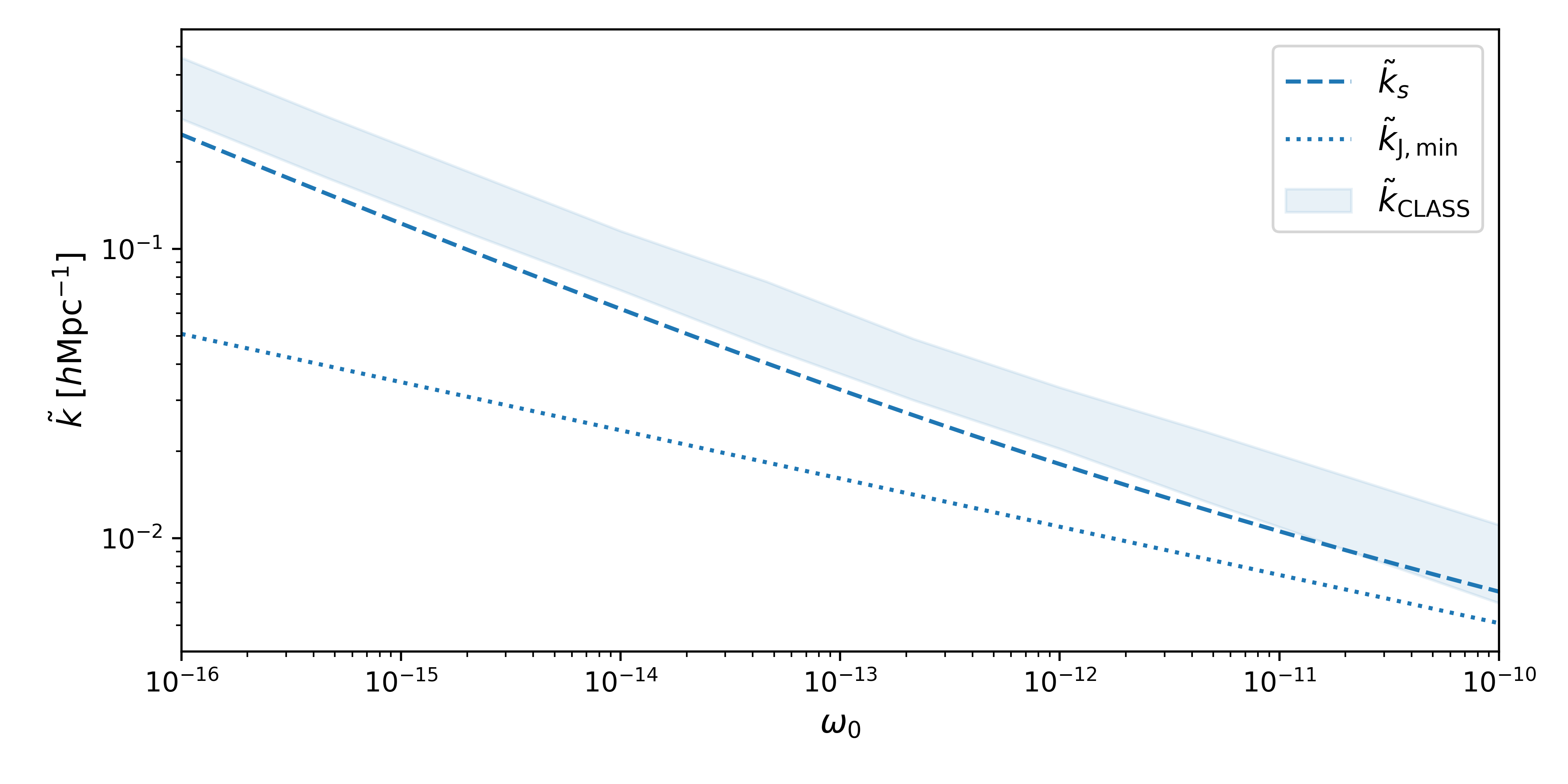}
    \includegraphics[width=0.7\linewidth]{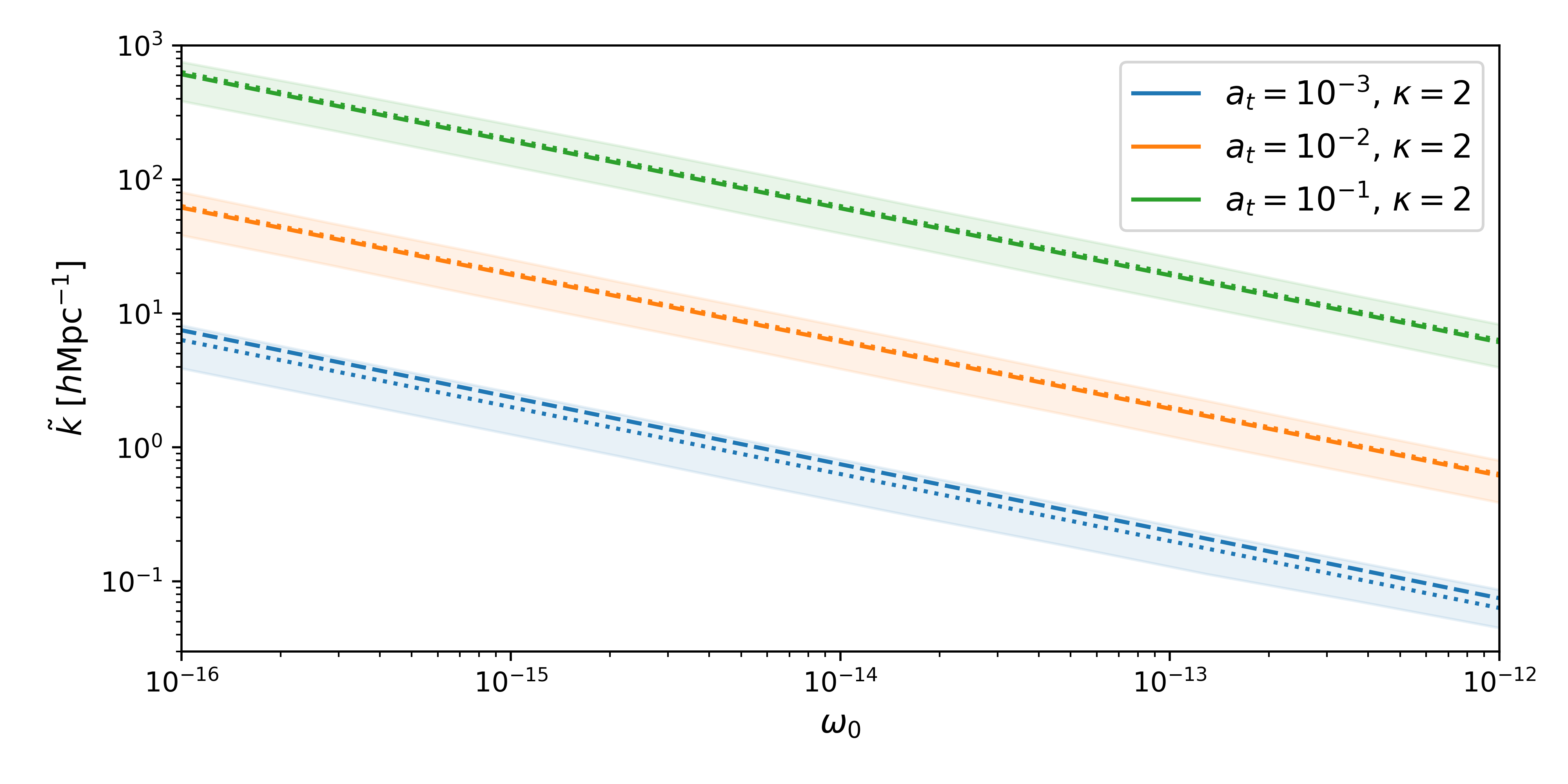}

    \caption{The comoving wavenumber $\Tilde{k}_{\text{CLASS}}$ (\textit{filled area}) where the matter power spectrum is $5\%$-$20\%$ lower than $\Lambda$CDM,  the minimum comoving Jeans' wavenumber $\Tilde{k}_{\text{J,min}}$ (\textit{dotted}), and the wavenumber $\Tilde{k}_{s}$ (\textit{dashed}) corresponding to the effective sound horizon, in a universe with pure SIBEC-DM (\textit{upper}) and a $\sigma_0$-SIBEC DM transition (\textit{lower}).}
\label{fig:k_Jeans_SIBEC}
\end{figure}

\begin{figure}
    \centering
    \includegraphics[width=0.7\linewidth]{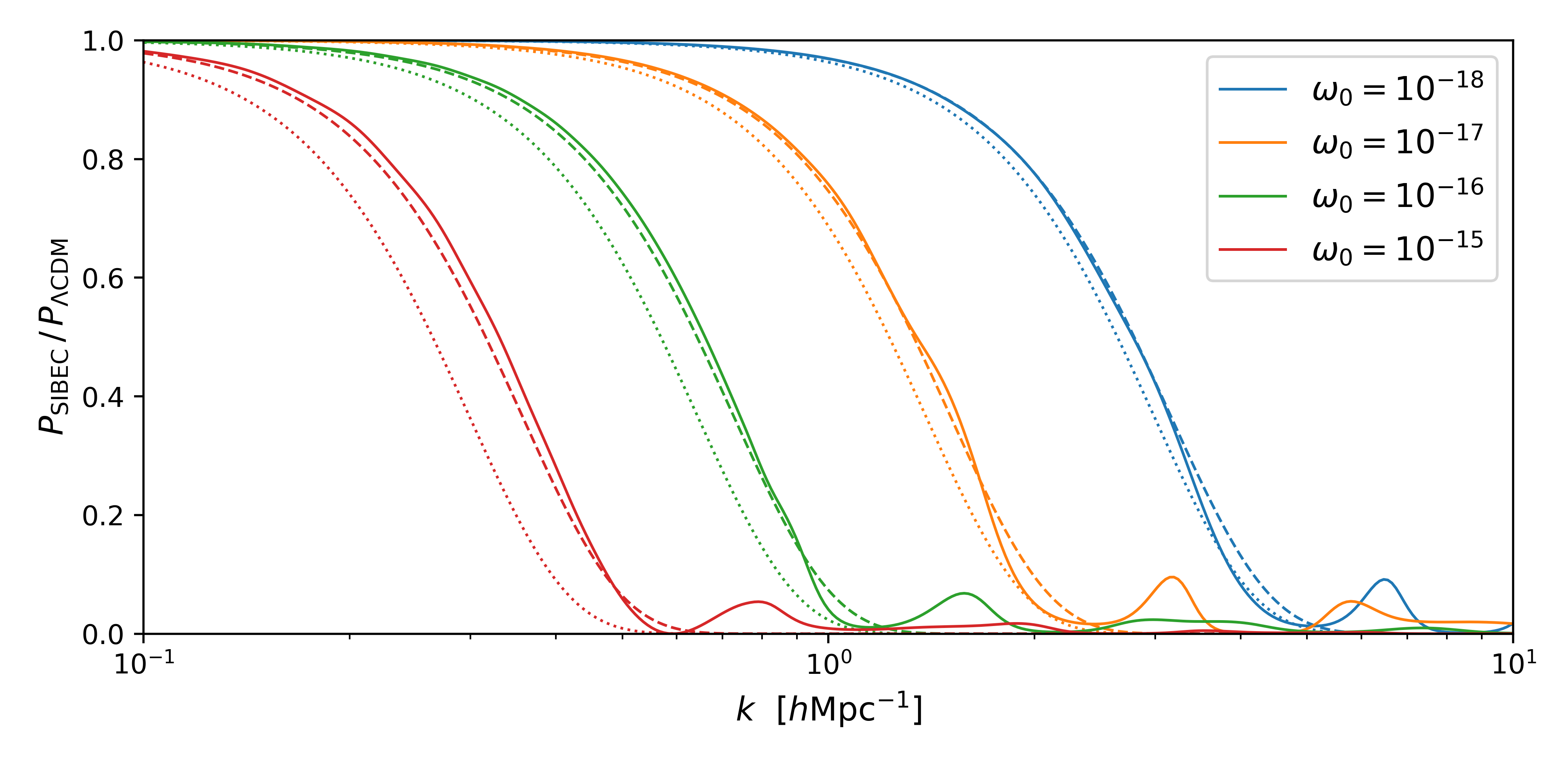}
    
    \caption{Suppression of the matter power spectrum in a SIBEC-DM universe compared to CDM (\textit{solid}), and the fitting function eq. \eqref{eq:Pk_suppression_approximation} with $\Tilde{k}_{\text{cut}}$ given by both the first equality (\textit{dashed}) and the final approximation (\textit{dotted}) in eq. \eqref{eq:kcut_pure_SIBEC}.}
\label{fig:pure_SIBEC_Pk_suppression}
\end{figure}

The primary effect of SIBEC-DM on the matter power spectrum, both with and without a transition, is a suppression of power below a certain scale due to a finite sound speed, and can be understood qualitatively from Jeans' instability. Considering the pure SIBEC-DM case for concreteness, the largest length-scale that resists gravitational collapse is given by the Jeans' wavelength
\begin{equation}
    \lambda_{\text{J}} = \frac{c_{s,\chi}}{\sqrt{G\rho_{\chi}}},
\end{equation}
where the sound speed and the density is evaluated at the background. Perturbations on scales below $\lambda_{\text{J}}$ are protected from gravitational instability by the fluid pressure and are "filtered" away, while scales far above are free to collapse and form structure in much the same way as CDM. To find which scales today that were affected by Jeans-filtering at some point in the past, the comoving Jeans' wavenumber is considered
\begin{equation}
    \Tilde{k}_{\text{J}} \approx \frac{a}{\lambda_{\text{J}}} = a\frac{\sqrt{G\rho_{\chi}}}{c_{s,\chi}}.
\end{equation}
The largest scale, or equivalently, the smallest wavenumber, that is subject to a significant suppression of power due to a finite fluid sound speed can therefore be approximated as the smallest Jeans' wavenumber $\Tilde{k}_{\text{J,min}}$ with respect to the scale factor. With the SIBEC sound speed, eq. \eqref{eq:SIBEC_cs2}, the smallest Jeans' wavenumber occurs around the time when SIBEC-DM becomes non-relativistic (which happens at $a\approx \omega_0^{1/3}$), and is roughly equal to
\begin{equation}
    \Tilde{k}_{\text{J,min}} \approx \frac{\sqrt{3G\rho_{0\chi}}}{\omega_0^{1/6}}.
\end{equation}
Another way to find the largest scales affected by the SIBEC-DM sound speed is given by the effective sound horizon
\begin{equation}
    r_{\text{s}} = \int_{0}^{\tau_0} \text{d}\tau \,c_{s,\chi} = \int_{0}^{\infty} \text{d}z \frac{c_{s,\chi}}{H}, 
\end{equation}
which corresponds to $\Tilde{k}_{\text{s}} \approx 1/r_{\text{s}}$. In figure~\ref{fig:k_Jeans_SIBEC}, both $\Tilde{k}_{\text{s}}$ and $\Tilde{k}_{\text{J,min}}$ are compared to $\Tilde{k}_{\mathrm{CLASS}}$, defined as the wavenumbers for which the linear matter power spectrum from CLASS is between $5\%$ and $20\%$ lower than in $\Lambda$CDM. The Jeans' length estimate $\Tilde{k}_{\text{J,min}}$ roughly reproduces the typical scale affected by pure SIBEC-DM, though with the wrong slope, while $\Tilde{k}_{\text{s}}$ is much more accurate. In fact, we find that the suppression of the matter power spectrum in a SIBEC-DM universe is to a good approximation
\begin{equation}
\label{eq:Pk_suppression_approximation}
    \frac{P_{\text{SIBEC}}}{P_{\Lambda\text{CDM}}} = e^{-(\Tilde{k}/\Tilde{k}_{\text{cut}})^3},
\end{equation}
where the cut-off scale $\Tilde{k}_{\text{cut}}$ is given by
\begin{equation}
\label{eq:kcut_pure_SIBEC}
    \Tilde{k}_{\text{cut}} = \frac{1}{3.2}\frac{2\pi}{r_s} \approx \frac{\sqrt{3\pi^3}}{8}\frac{H_0\sqrt{\Omega_{r0}}}{(6\omega_0)^{1/3}} \approx 0.2 h\text{Mpc}^{-1} \Bigg(\frac{R_c}{1\text{kpc}}\Bigg)^{-2/3},
\end{equation}
as shown in figure~\ref{fig:pure_SIBEC_Pk_suppression}. The second (approximate) equality is valid for $\omega_0$ such that SIBEC-DM becomes non-relativistic during radiation-domination, i.e. $\omega_0 \ll 10^{-12}$, or $R_c \ll 30\text{kpc}$. Eq. \eqref{eq:kcut_pure_SIBEC} is similar to the result obtained in ref. \cite{Shapiro2021} by considering the first $k$-mode that enters the horizon while it is sub-Jeans, though they found a slightly smaller exponent of $-1/2$ instead of $-2/3$. We find that the exponent in eq. \eqref{eq:kcut_pure_SIBEC} fits the results from CLASS better.

\begin{figure}
    \centering  
    \includegraphics[width=0.49\linewidth]{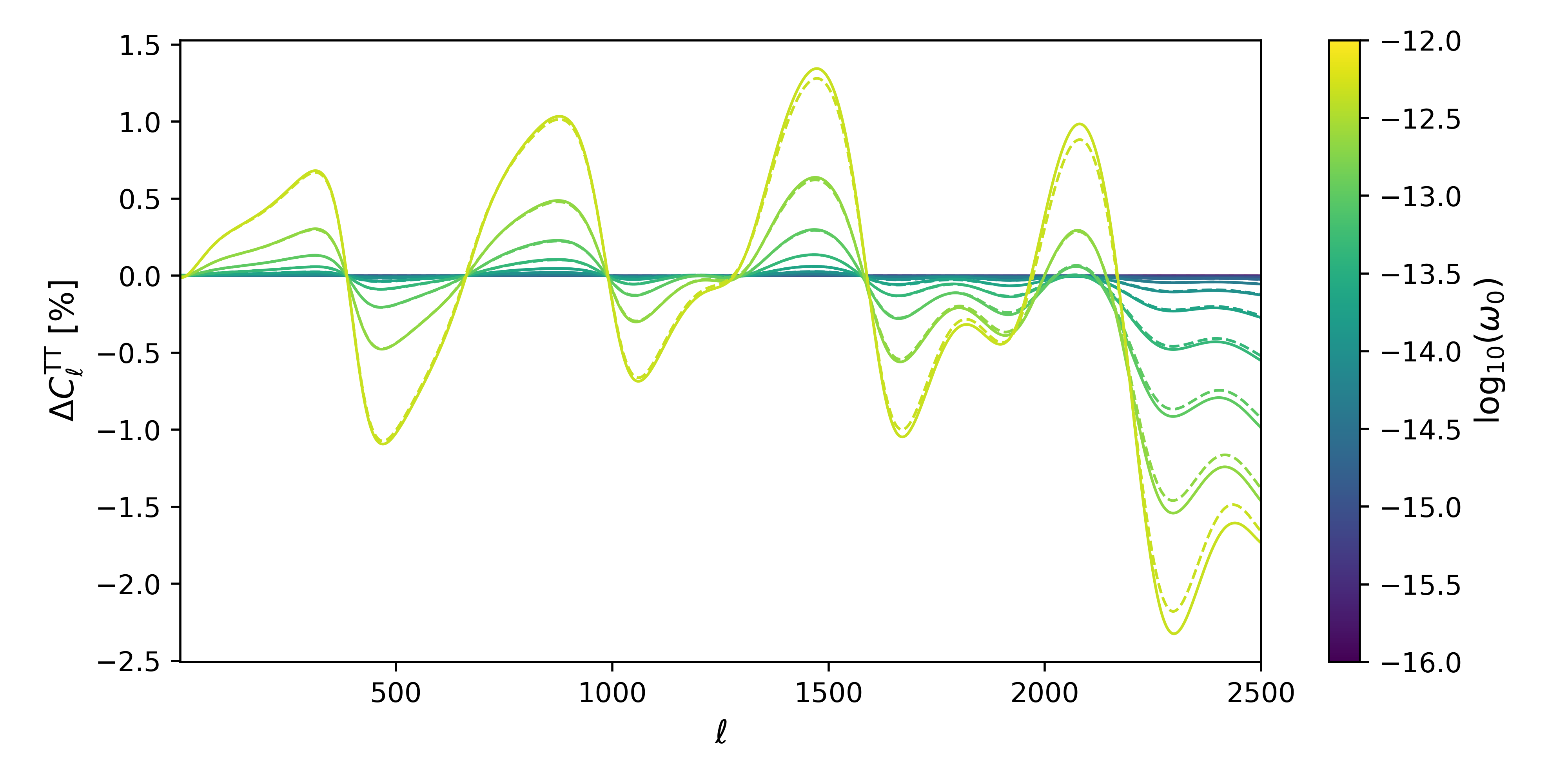}
    \includegraphics[width=0.49\linewidth]{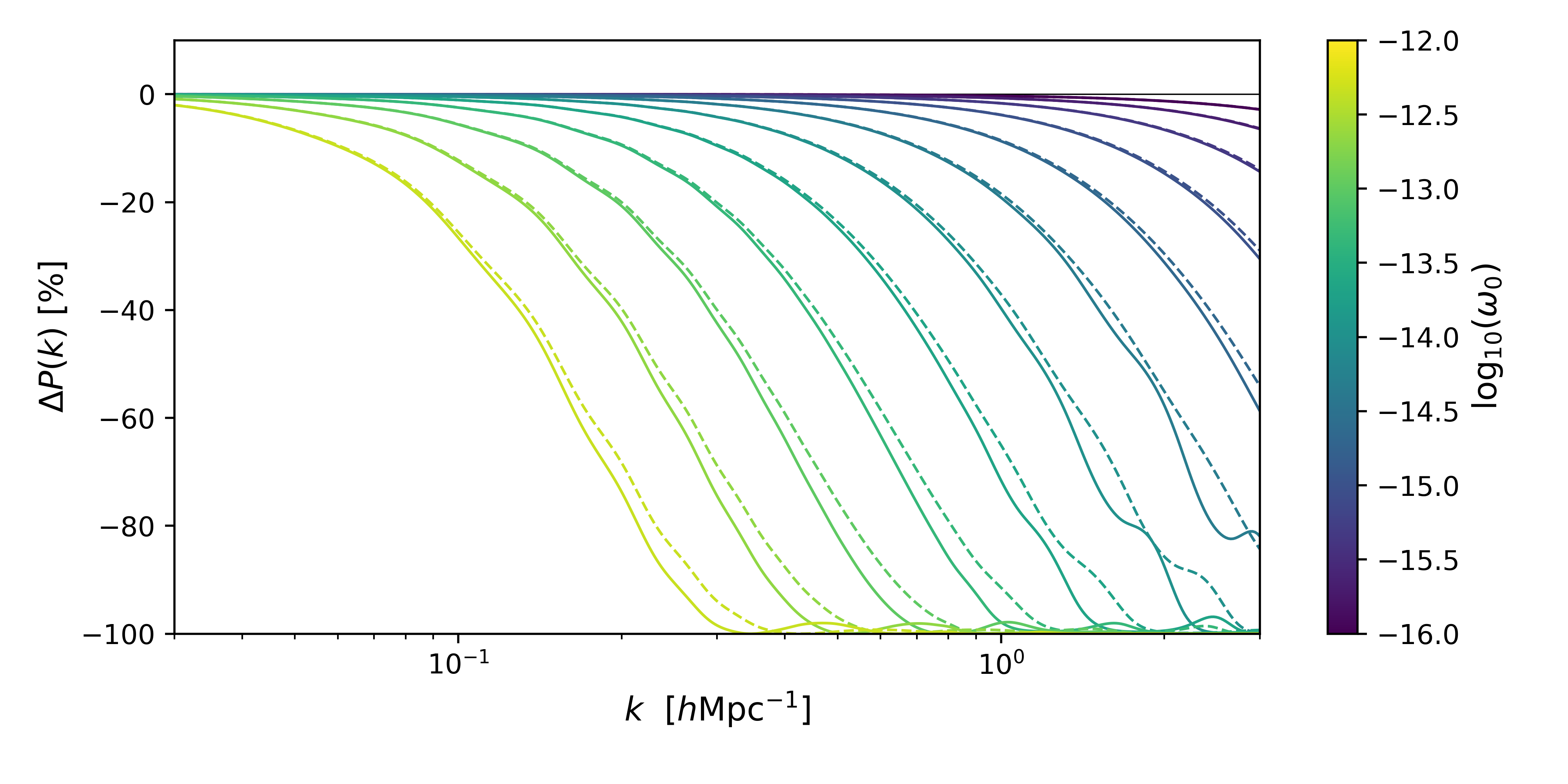}
    \includegraphics[width=0.49\linewidth]{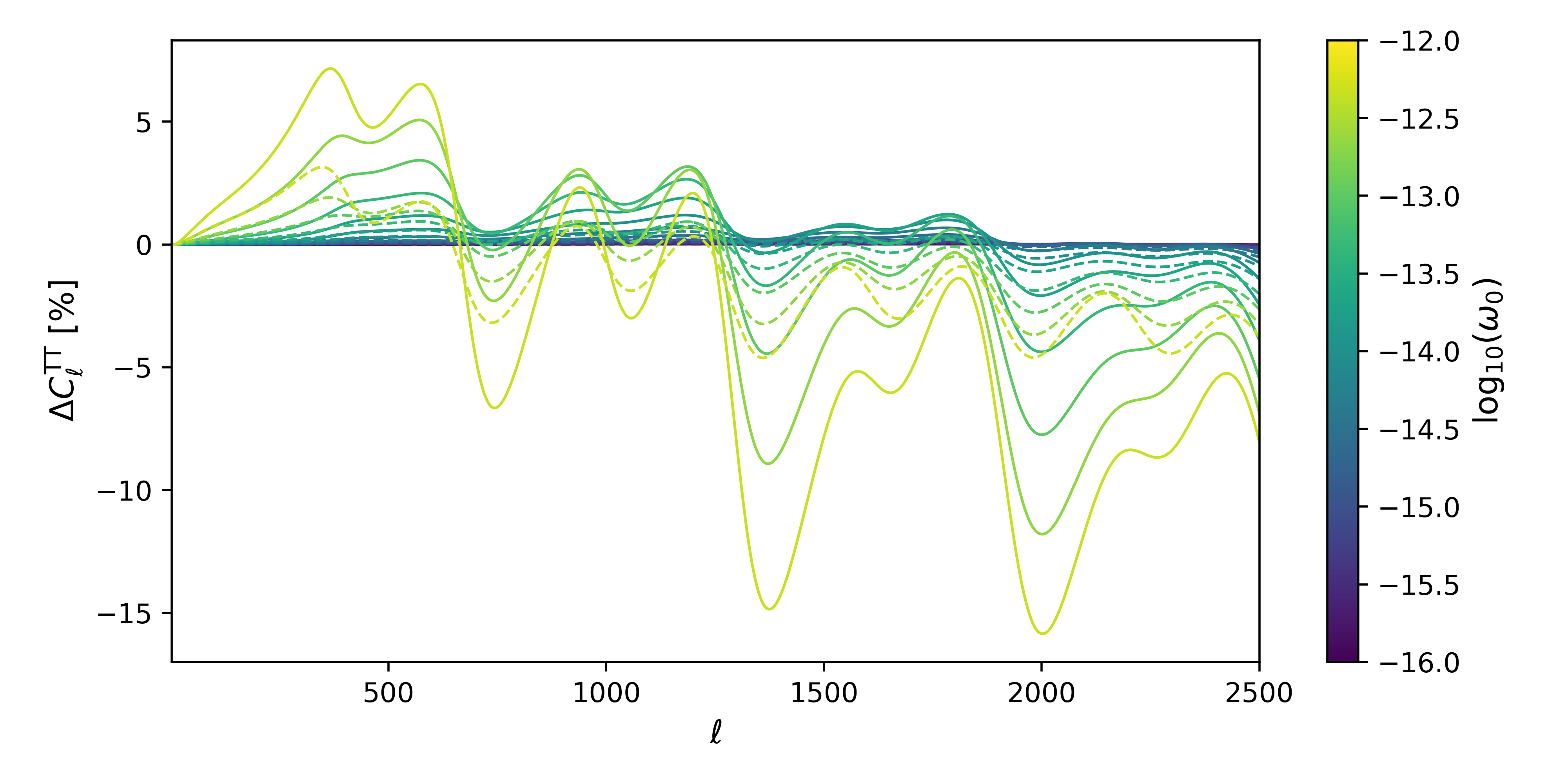}
    \includegraphics[width=0.49\linewidth]{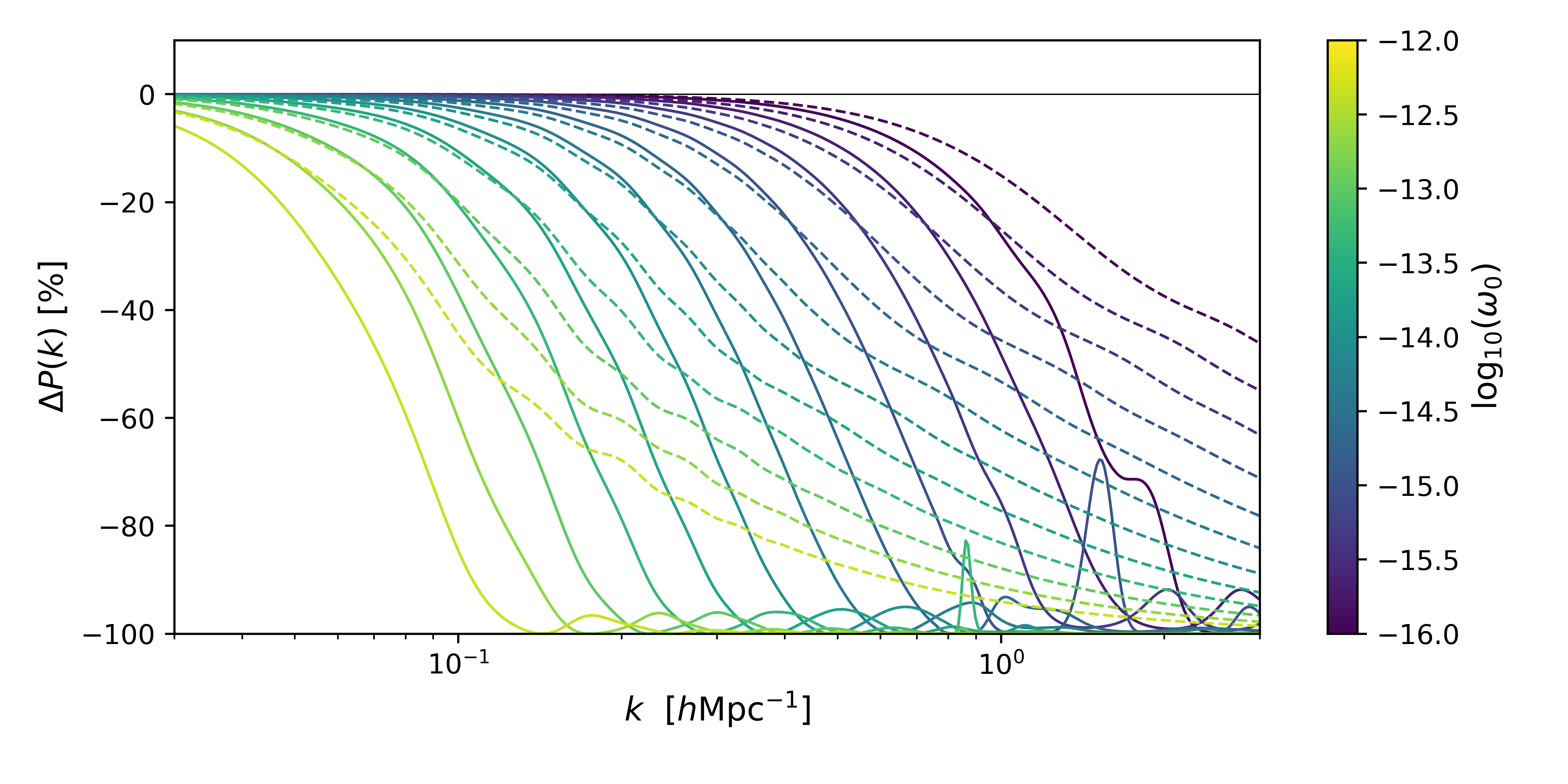}

    \caption{Deviation of CMB temperature and matter power spectra from $\Lambda$CDM in a cold-SIBEC-DM universe, using the single-fluid approach (\textit{solid lines}) and the two-fluid approach (\textit{dashed lines}). The upper plots have $\kappa=4$, while the lower have $\kappa=1$, with $a_t=10^{-3}$ throughout.}
\label{fig:cold_SIBEC_Pk_and_CTTl_kappa_1vs2_fluid}
\end{figure}

A similar estimate can be made for e.g. $\sigma_0$-SIBEC. Using the effective single-fluid sound speed eq. \eqref{eq:eq:constant_w_SIBEC_cs2} and assuming the energy density to be matter-like, i.e. that $\omega \ll 1/3$ at all times, gives the scale factor of $\Tilde{k}_{\text{J,min}}$ as $a_{\text{J,min}} \approx a_t$, such that 
\begin{equation}
    \Tilde{k}_{\text{J,min}} = a_t\sqrt{\frac{G\rho_{\chi 0}}{\omega_0}}.
\end{equation}
This result is not very dependent on $\kappa$, and both $\Tilde{k}_{s}$ and $\Tilde{k}_{\text{J,min}}$ agrees well with $\Tilde{k}_{\text{CLASS}}$, also shown in figure~\ref{fig:k_Jeans_SIBEC}. In fact, the effect of the parameterization in eq. \eqref{eq:eq:constant_w_SIBEC_eos} on the matter and CMB power spectrum in the single-fluid approach does not depend much on $\kappa$. The two-fluid approach, on the other hand, is more sensitive to $\kappa$, though it matches closely to the single-fluid for large $\kappa$, i.e it behaves as a single fluid with an effective equation of state. This observation is easy to understand intuitively; instantaneously replacing one component, including all of its perturbations, for another component with a different equation of state is essentially equivalent to suddenly changing the equation of state from one phase to the other. For slow transitions the two approaches generally differ. For $\sigma_0$-SIBEC-DM the two-fluid case results in a stronger suppression of power, as shown in figure~\ref{fig:cold_SIBEC_Pk_and_CTTl_kappa_1vs2_fluid}, since after the SIBEC-DM transition there is a sizable portion of the initial component still present that has a larger sound speed than the SIBEC-DM component. In the case of a CDM-like initial phase, however, the opposite is true; the two-fluid approach gives a softer damping of the matter power spectrum and a smaller deviation from $\Lambda$CDM in the CMB for slow transitions, as shown in figure~\ref{fig:cold_SIBEC_Pk_and_CTTl_kappa_1vs2_fluid}, since throughout most of the cold-SIBEC DM transition, also for a time after $a_t$, there remains some CDM that drives the perturbations of the emerging SIBEC-DM component towards $\Lambda$CDM, whereas in the single-fluid approach all of DM is essentially SIBEC-like after $a_t$, regardless of the rate of the transition.

\begin{figure}
    \centering  
    \includegraphics[width=0.49\linewidth]{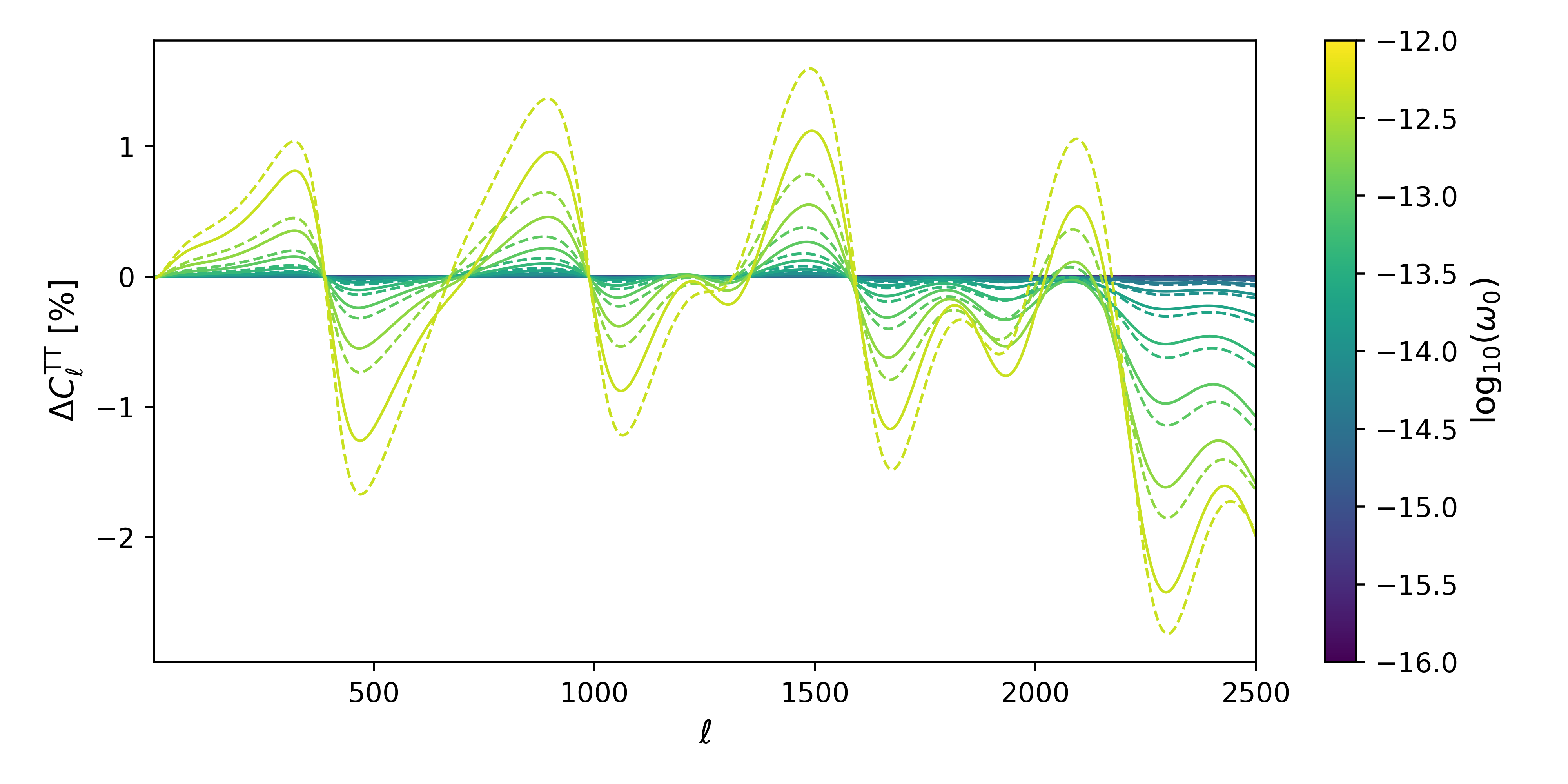}
    \includegraphics[width=0.49\linewidth]{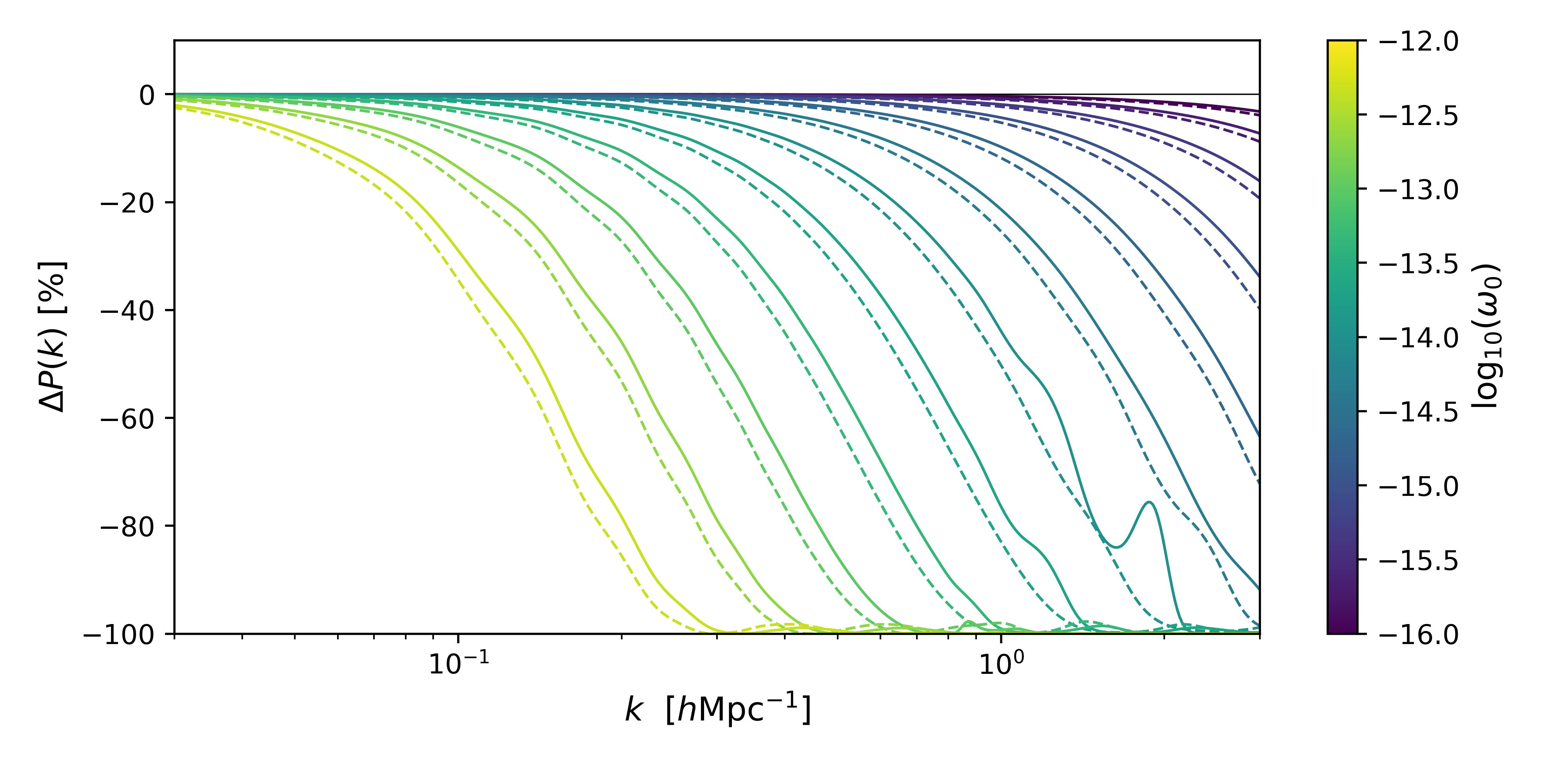}
    \includegraphics[width=0.49\linewidth]{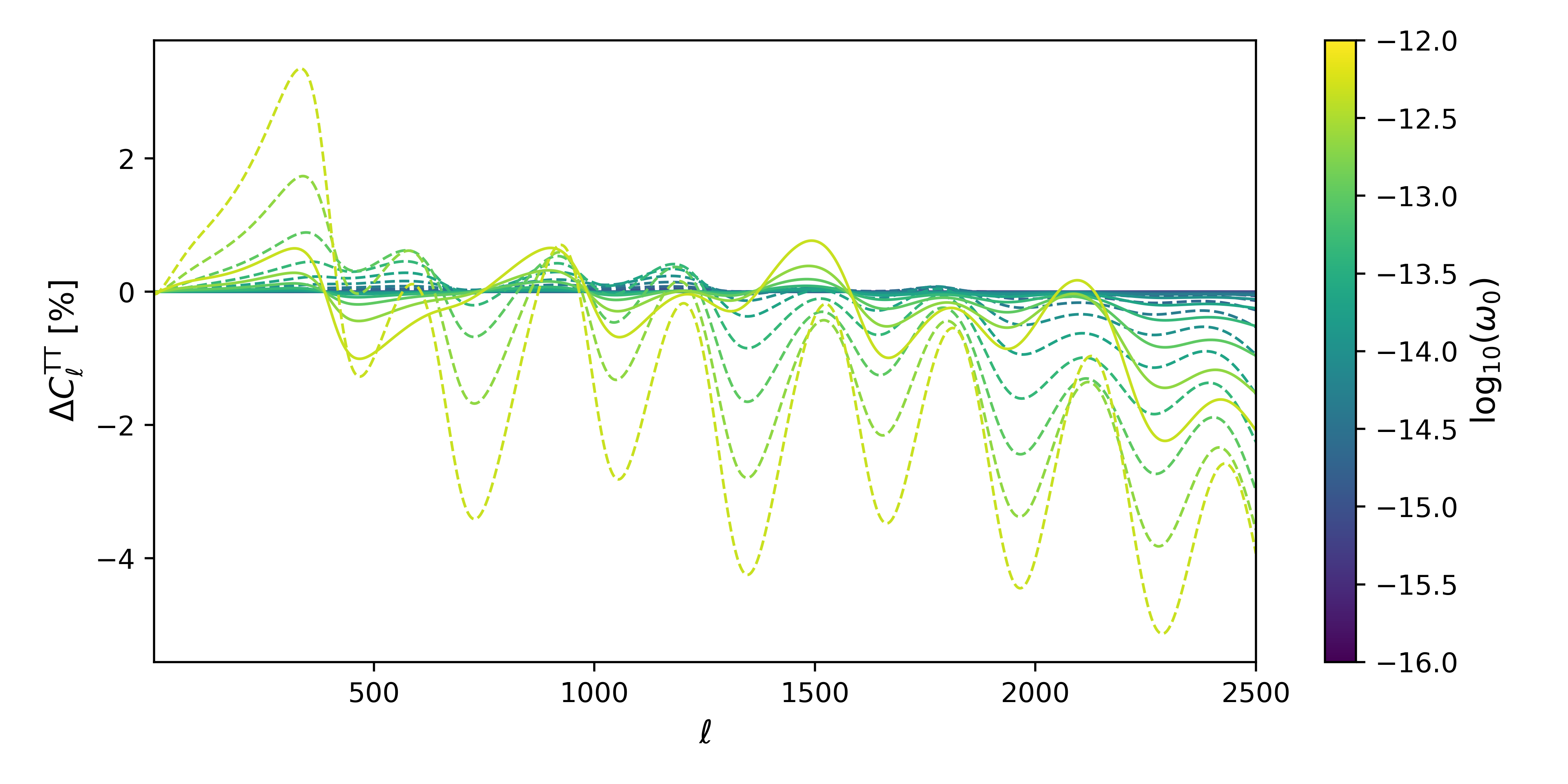}
    \includegraphics[width=0.49\linewidth]{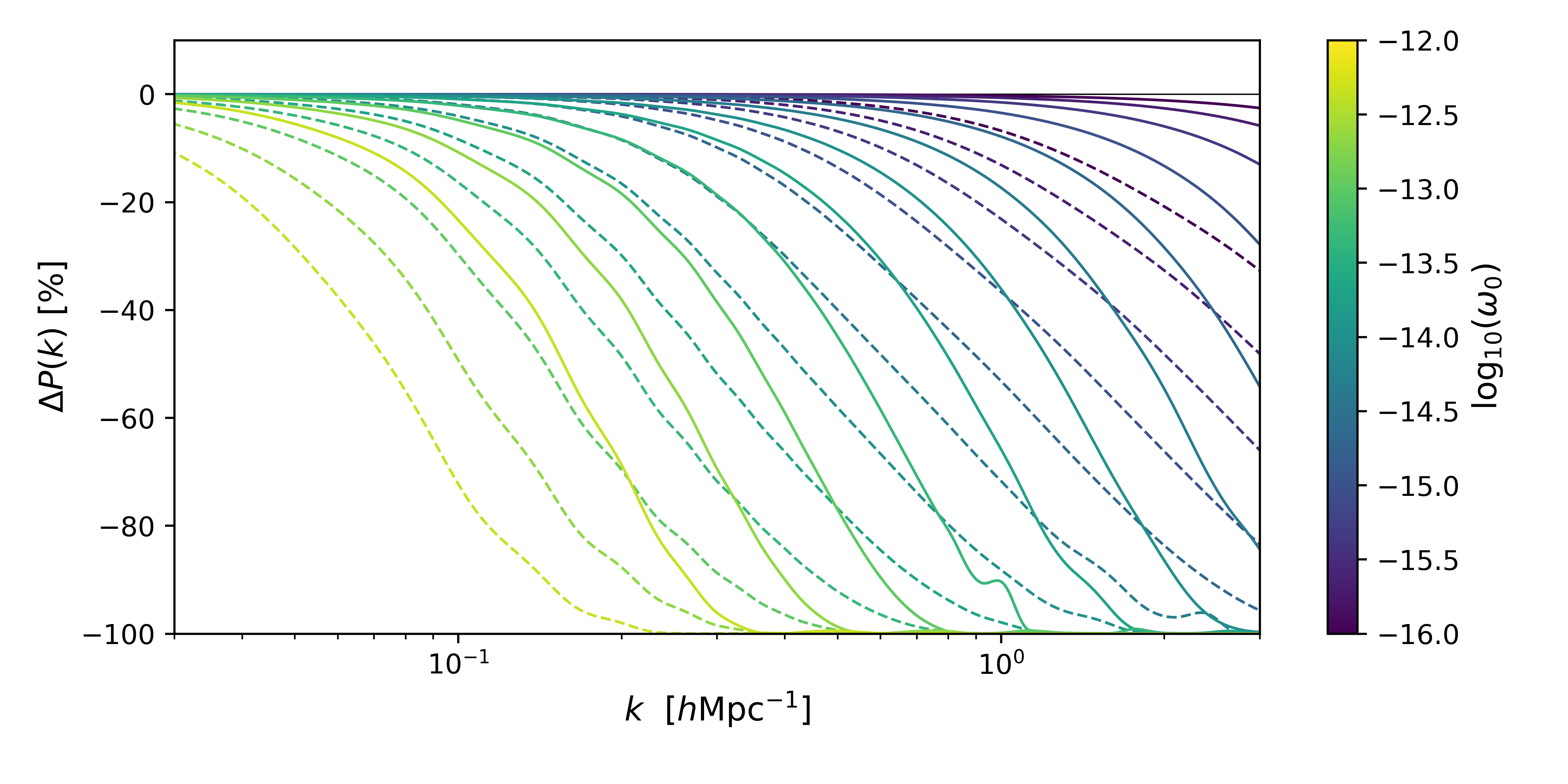}

    \caption{Deviation of CMB temperature and matter power spectra from $\Lambda$CDM in a $\sigma_0$-SIBEC DM universe, using the single-fluid approach (\textit{solid lines}) and the two-fluid approach (\textit{dashed lines}). The upper plots have $\kappa=6$, while the lower have $\kappa=1$, with $a_t=10^{-3}$ throughout.}
\label{fig:constant_w_SIBEC_Pk_and_CTTl_kappa_1vs2_fluid}
\end{figure}

Altering the phenomenology of DM also affects the CMB, and one of its main features is the position of the acoustic peaks, which are, in a flat universe, determined by the sound horizon $r_*$ of the baryon-photon fluid at recombination $z_*$, and the distance $D_*$ to the surface of last scattering. The angular size of the sound horizon $\theta_*$ (which corresponds to the first peak in the CMB) is given by
\begin{equation}
    \theta_* = \frac{r_*}{D_*},
\end{equation}
where
\begin{equation}
    r_* = \frac{1}{\sqrt{3}}\int_{z_*}^{\infty}\frac{\text{d}z}{H\sqrt{1+R}},
\end{equation}
\begin{equation}
    D_* = \int_{0}^{z_*}\frac{\text{d}z}{H},
\end{equation}
with $R=a3\Omega_{\text{b}0}/4\Omega_{\gamma 0}$ \cite{Dodelson2003}. Since only dark matter is modified, with the rest of $\Lambda$CDM left unchanged, $r_*$ and $D_*$ are only affected by SIBEC-DM through a change in the Hubble parameter; an increase in $H$ prior to recombination shrinks the sound horizon, and an increase in $H$ between recombination and today instead shrinks $D_*$. Since SIBEC-DM is matter-like at late times, and the change in $H$ is largest at early times, then $D_*$ is largely the same as in $\Lambda$CDM, while $r_*$ shrinks. There is therefore a reduction in $\theta_*$, and the acoustic peaks are shifted towards smaller angular scales.

SIBEC-DM also affects the height of the peaks through e.g. the ratio of matter to radiation, which moves the era of matter-radiation equality, since the observed CMB temperature of modes that enter the horizon during radiation-domination experiences a boost as the gravitational potential decays and drives the acoustic oscillations. Furthermore, potential decay due to a finite sound speed after recombination contributes to the integrated Sachs-Wolfe effect between recombination and today, and reduces the lensing signal in the CMB. A detailed discussion of how the CMB is affected by the parameters of the GDM framework can be found in refs.  \cite{Hu1998, Kopp2016, Thomas2016}.

\section{Constraints from cosmological data}
\label{sec:constrains_from_cosmological_data}
Constraints are placed on SIBEC-DM with CMB temperature and polarization power spectra from the Planck 2018 data release \cite{Planck2020}, BAO and growth rate measurements from BOSS \cite{Alam2017}, and the Pantheon compilation of SNIa distances \cite{Scolnic2018}, using the MCMC code MontePython \cite{Audren2012,Brinckmann2018}, and the Boltzmann code CLASS for computing the observables \cite{Blas2011}, modified to include the fluid models of DM described in this work at the background and linear level. The priors in the MCMC runs are listed in table \ref{tab:priors}, and the MCMC chains were checked using Gelman-Rubin $R$-statistics \cite{Gelman1992}, with convergence assumed when $R-1<0.01$ for all the parameters.

\bgroup 
\def\arraystretch{1.3}
\begin{table}[t]
\caption{Priors on the cosmological parameters varied in the MCMC runs, not including the nuisance parameters for the CMB and Pantheon datasets.\newline}              
\label{tab:priors}      
\centering                                      
\begin{tabular}{l l l l}          
\hline\hline
Model & Parameter  & Symbol               & Prior \\
\hline
All & DM density today                  & $\Omega_{\text{dm}0}h^2$ & [0.05, 0.3]  \\
    & Baryon density today              & $\Omega_{\text{b}0}h^2$  & [0.01, 0.05] \\
    & Amplitude of scalar spectrum      & $\ln(10^{10}A_s)$        & [2, 4]       \\
    & Scalar spectral index             & $n_s$                    & [0.8, 1.1]   \\
    & Reionization optical depth                     & $\tau_{\text{reio}}$                   & [0.01, 0.1]  \\
    & Angular size of the sound horizon & $100\theta_*$            & [0.8, 1.2]   \\
\hline
pure SIBEC-DM (\#1)      & SIBEC equation of state today       & $\log_{10}(\omega_0)$    & [-18, -12]   \\
\hline
pure SIBEC-DM (\#2)      & SIBEC equation of state today       & $\omega_0$    & [0, $10^{-12}$]   \\
\hline
$\sigma_0$-SIBEC-DM & Initial constant equation of state  & $\log_{10}(\sigma_0^2)$  & [-10, -2]   \\
                    & Scale factor at transition          & $\log_{10}(a_t)$         & [-5, 0]      \\
\hline
warm-SIBEC-DM       & SIBEC equation of state today       & $\log_{10}(\omega_0)$    & [-18, -12]   \\
                    & Scale factor at transition          & $\log_{10}(a_t)$         & [-6, 0]      \\
\hline
cold-SIBEC-DM       & SIBEC equation of state today       & $\log_{10}(\omega_0)$    & [-18, -10]   \\
                    & Scale factor at transition          & $\log_{10}(a_t)$         & [-6, 0]      \\
\hline\hline
\end{tabular}
\end{table}
\egroup 

\begin{figure}
    \centering
    \includegraphics[width=0.49\linewidth]{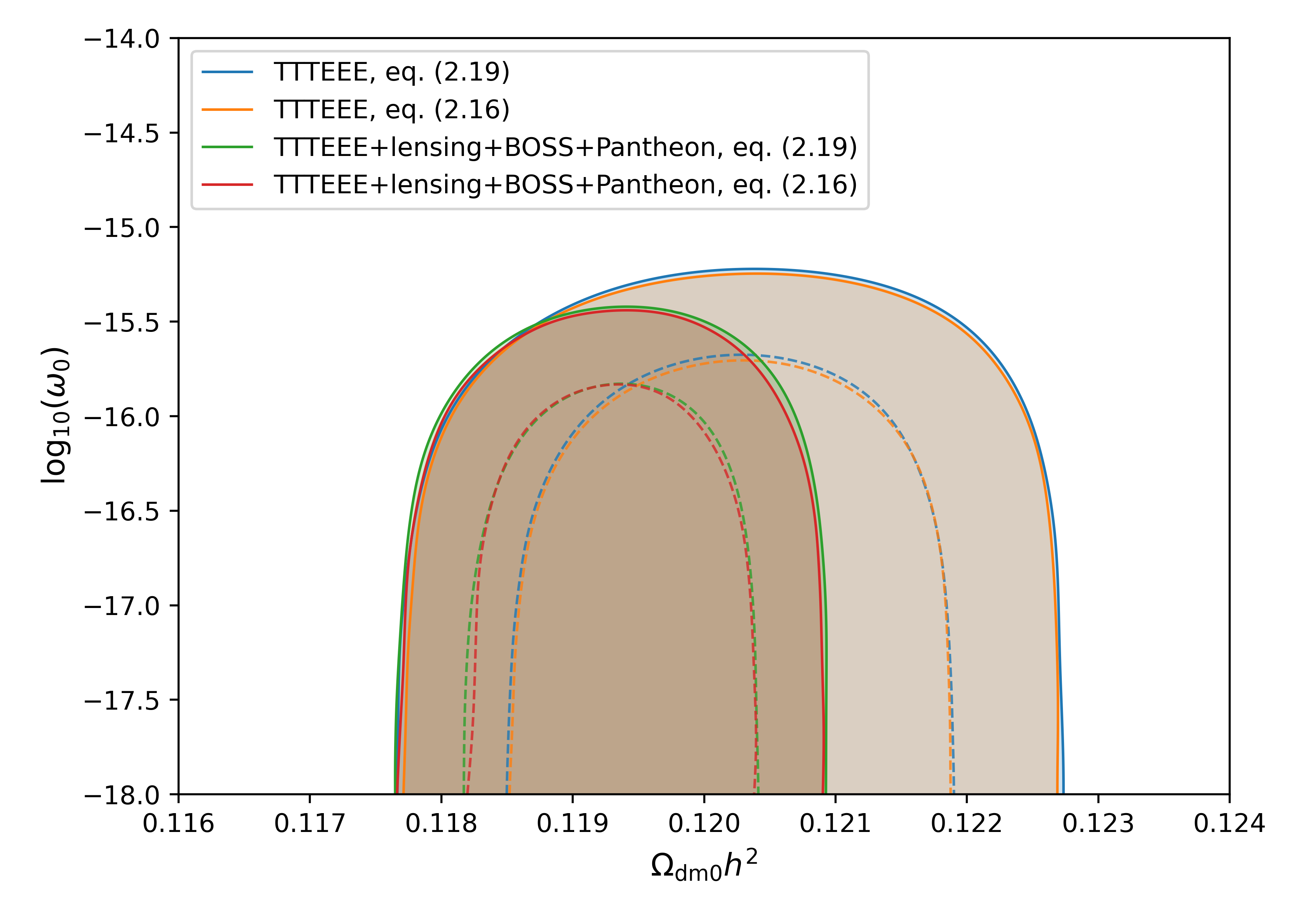}
    \includegraphics[width=0.49\linewidth]{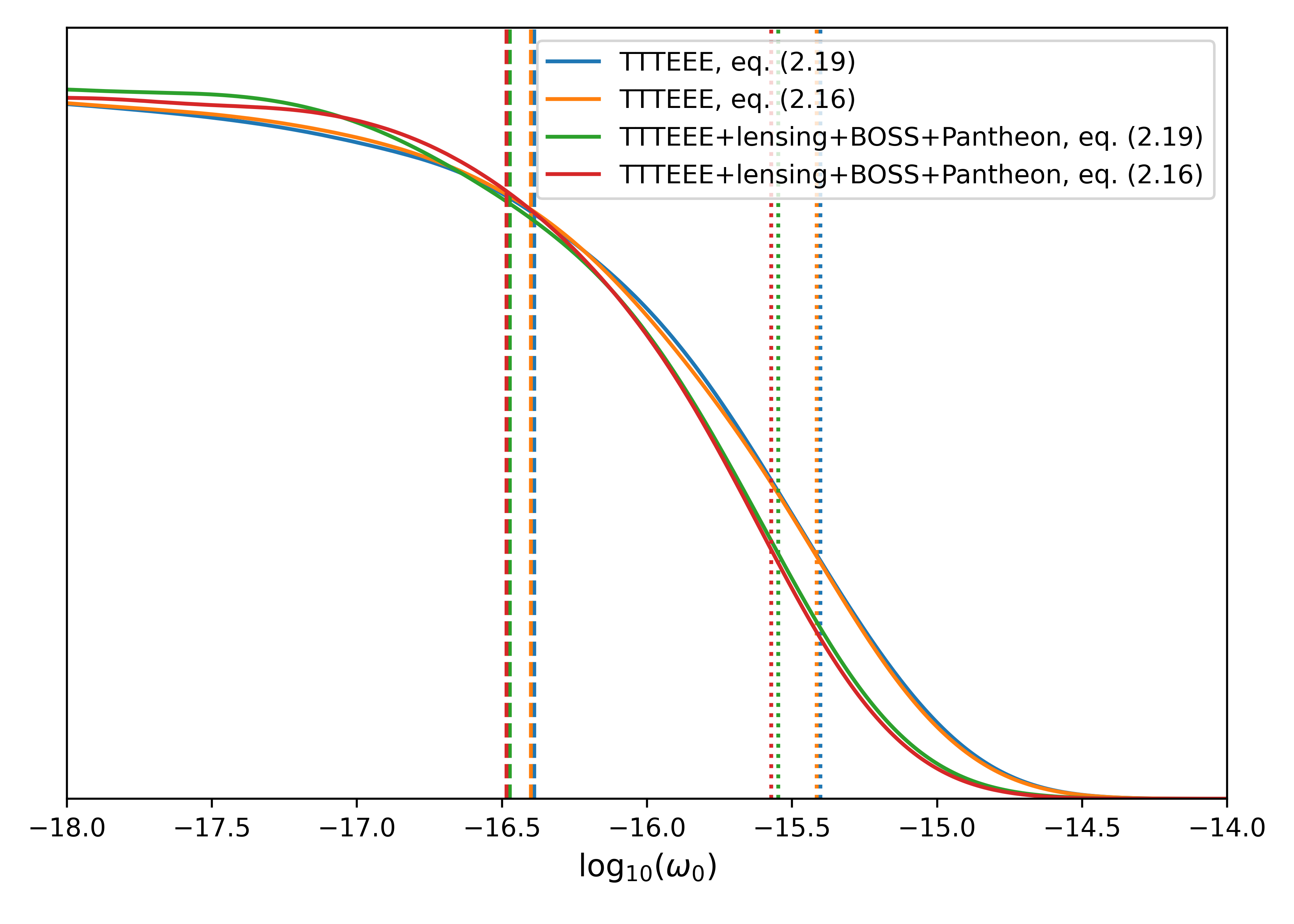}
    \includegraphics[width=0.49\linewidth]{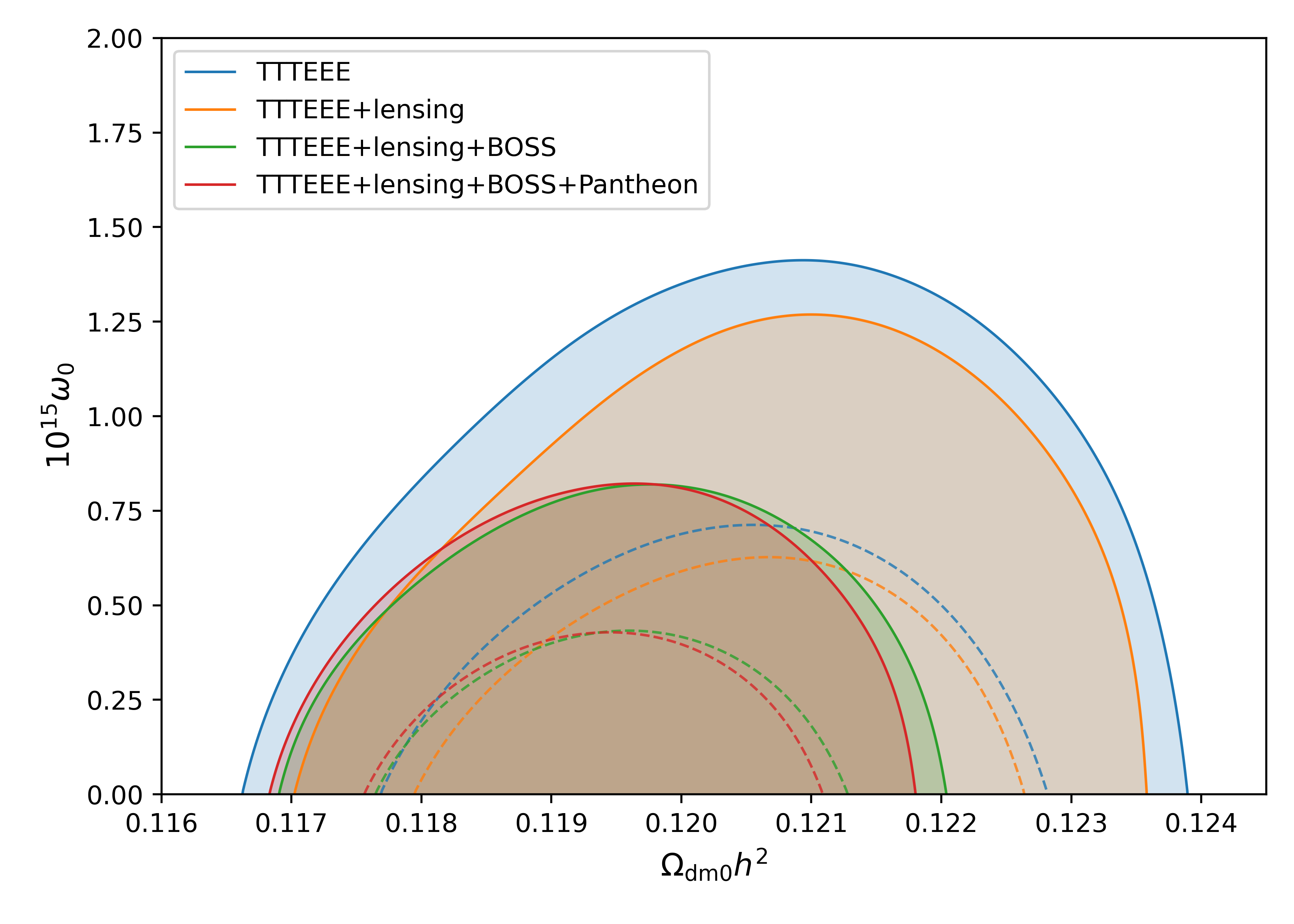}
    \includegraphics[width=0.49\linewidth]{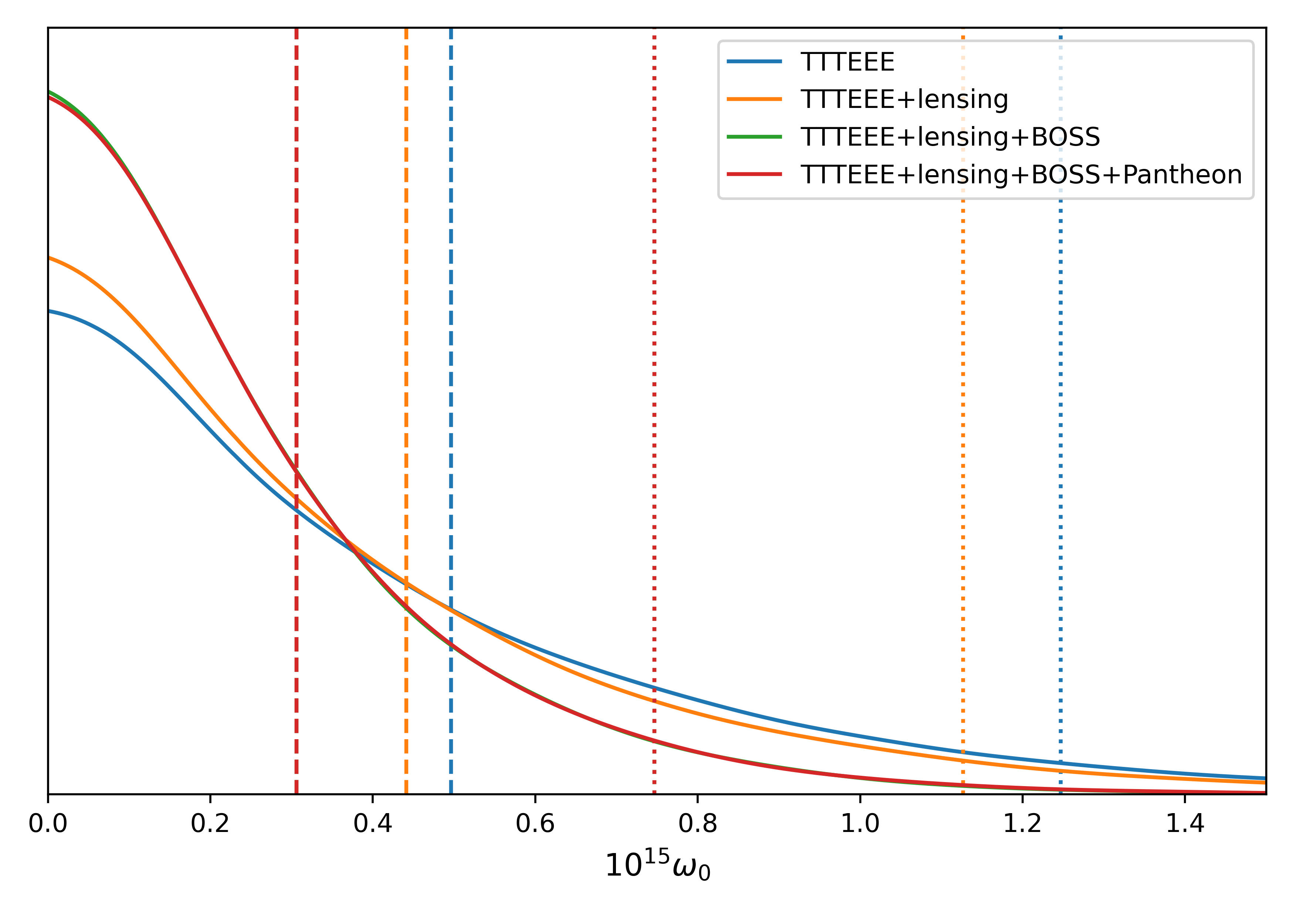}

    \caption{Credible regions for $\Omega_{\text{dm0}}h^2$ and $\omega_{0}$ for pure SIBEC-DM. The 95\% (\textit{shaded contours}) and 68\% (\textit{colored dashed lines}) contours are shown on the left, and the 1D posterior for $\omega_0$, with the $68\%$ and $95\%$ bounds shown with dashed and dotted vertical lines, respectively, on the right.
    The upper row compares the constraints obtained from the two approximations for the SIBEC-DM equation of state eqs. \eqref{eq:SIBEC_eos} and \eqref{eq:SIBEC_eos_improved} for a flat prior on $\log_{10}(\omega_0)$. The lower row shows the constraints for a flat prior on $\omega_0$ using eq. \eqref{eq:SIBEC_eos_improved}.}
\label{fig:mcmc_pure_SIBEC}
\end{figure}

Bounds on the self-interaction of pure SIBEC-DM are shown in figure~\ref{fig:mcmc_pure_SIBEC} for flat priors on $\log_{10}(\omega_0)$, as well as a comparison of the two SIBEC-DM equation of state eqs. \eqref{eq:SIBEC_eos} and \eqref{eq:SIBEC_eos_improved}, which shows a negligible difference between the two approximations. As increasingly negative values for $\log_{10}(\omega_0)$ are favoured by the data (corresponding to the CDM-limit of SIBEC-DM), we find that the constraints on $\log_{10}(\omega_0)$ are sensitive on the prior range. This is a feature of any logarithmic parameter that is uniformly sampled without a well-motivated boundary (see e.g. section VIII C in ref. \cite{Joudaki2020} for a very similar case). In our case, the limit $\omega_0\rightarrow 0$ can be made arbitrarily large in log-space, and therefore result in overly restrictive bounds. For instance, with a lower bound of $\log_{10}(\omega_0) > -18$ to the prior range we get $\omega_0<3.3\times 10^{-17}$ at $68\%$ confidence and $\omega_0<2.7\times 10^{-16}$ at $95\%$, but if we instead place the lower bound at $\log_{10}(\omega_0) > -20$ we get $\omega_0<6.6\times 10^{-18}$ at $68\%$ and $\omega_0<1.7\times 10^{-16}$ at $95\%$, while $\log_{10}(\omega_0) > -16$ gives $\omega_0<2.5\times 10^{-16}$ at $68\%$ and $\omega_0<6.2\times 10^{-16}$ at $95\%$. We have for this reason also included results of chains where we impose a flat prior on $\omega_0$ instead, also shown in figure~\ref{fig:mcmc_pure_SIBEC}. This provides weaker constraints on the self-interaction of SIBEC-DM, but does not suffer from the issue of a possibly ill-chosen range on the parameter that is varied during the MCMC runs. We quote the constraints from these chains in the following for pure SIBEC-DM. In the cases where transitions are included the bounds on $\omega_0$ span several orders in magnitude as $a_t$ is varied, hence only flat priors on $\log_{10}(\omega_0)$ is used here. It is therefore important to keep in mind that these constraints are dependent on the prior range, the $68\%$ confidence region more so than the $95\%$ region.

\begin{figure}
    \centering 
    \includegraphics[width=0.99\linewidth]{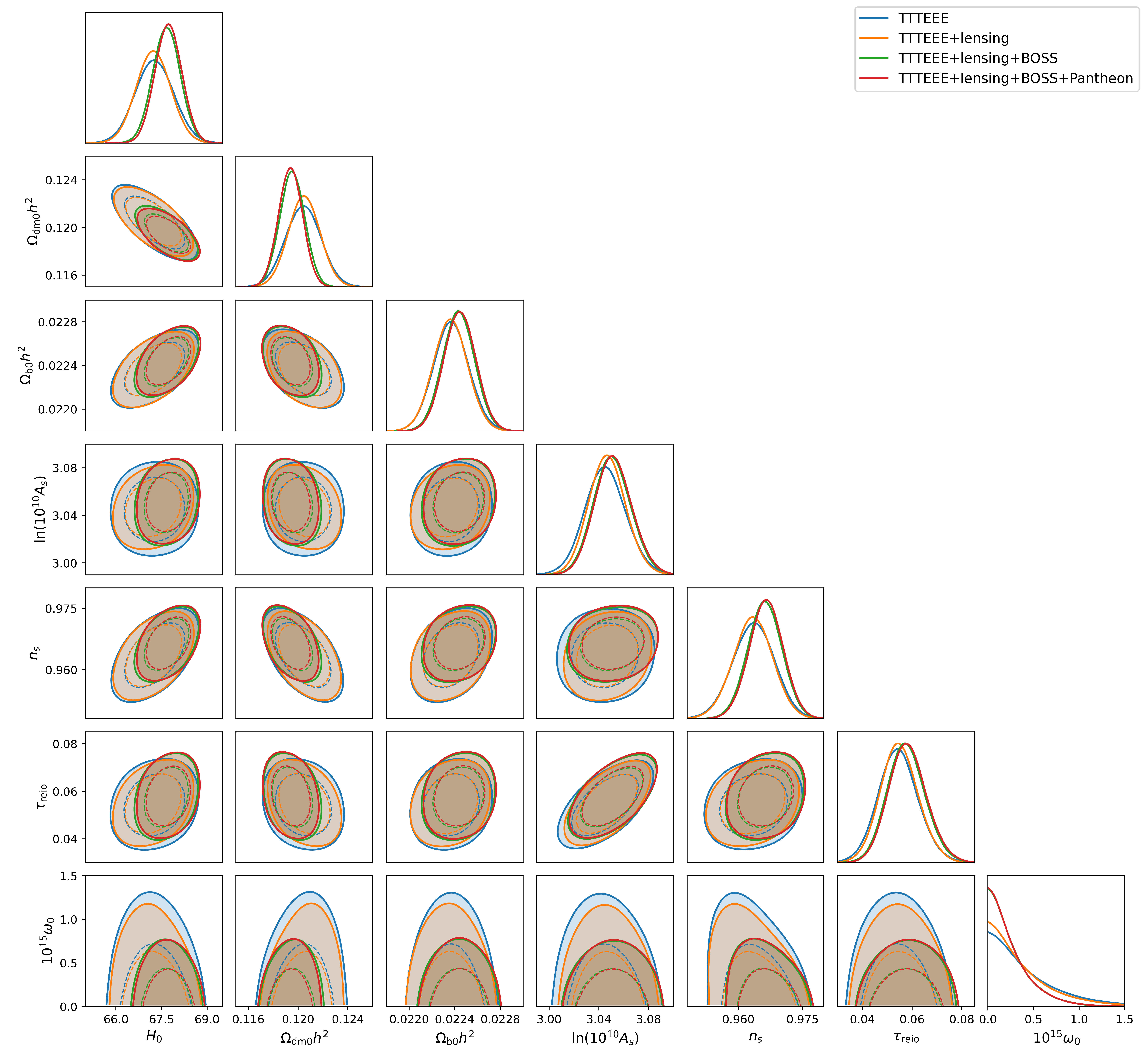}

    \caption{The $95\%$ (\textit{colored contours}) and $68\%$ (\textit{colored dashed lines}) confidence regions for the standard cosmological parameters and pure SIBEC-DM, as well as 1D posteriors, with a flat prior on $\omega_0$.}
\label{fig:mcmc_triangle_pure_SIBEC}
\end{figure}

We find that $\omega_0<3.0\times10^{-16}$ at $1\sigma$ for pure SIBEC-DM, and that the self-interaction does not show any strong degeneracies with the standard cosmological parameters, whose posteriors are largely unchanged from $\Lambda$CDM, shown in figure~\ref{fig:mcmc_triangle_pure_SIBEC}. This can be compared to the upper limits on $\omega_0$ obtained from the background evolution of SIBEC-DM (which are also valid for scenarios with very early transitions). For instance, matter-radiation equality $z_{\text{eq}}$ should be within the Planck 2018 range \cite{Planck2020}, i.e. not deviate from $\Lambda$CDM by more than around $0.6\%$, which gives $\omega_0 \lesssim 3.5\times10^{-14}$. An additional bound can be found by demanding that the effective number of relativistic degrees of freedom, or effective number of neutrino species $N_{\text{eff}}$, not be larger than that inferred from the observed abundance of light elements. Following ref. \cite{Li2014}, the change in $N_{\text{eff}}$ due to the SIBEC-component, $\Delta N_{\text{eff, }\chi}$, which at the time of BBN is entirely radiation-like for the values of $\omega_0$ of interest, is defined as
\begin{equation}
    \frac{\Delta N_{\text{eff, }\chi}}{N_{\text{eff, standard}}} = \frac{\bar{\rho}_{\chi}}{\bar{\rho}_{\nu}}.
\end{equation}
The standard value for the effective number of neutrino species in $\Lambda$CDM is $N_{\text{eff, standard}} = 3.046$. Hence, if the contribution of SIBEC-DM to dark radiation at around BBN is limited to $\Delta N_{\text{eff, }\chi} \lesssim 0.5$ \cite{Steigman2012,Nollett2015}, then $\omega_0 \lesssim 4\times 10^{-15}$.\footnote{Note that this estimate, as in the rest of our analysis, does not include the very early stiff phase in ref. \cite{Li2014}, which can also affect BBN.} The upper bound on $\omega_0$ in this work is therefore between one and two orders in magnitude lower than those obtained from background considerations. This is of particular interest with regards to the core-cusp problem. At the very least, the core radii of SIBEC-DM halos must not be much larger than the smallest observed structures if SIBEC-DM is the dominant DM-component today. Since the radius of these halos is related to the self-interaction of SIBEC-DM, then a condition can be obtained by using eq. \eqref{eq:w0_limit_hydrostatic_equilibrium} with e.g. $R_c < 10\text{kpc}$ as an upper limit on the typical core radius, giving $\omega_0 \lesssim 10^{-13}$. For SIBEC-DM to be a possible solution to the cusp-core problem, then $R_c$ should instead not be much smaller than the typical core-size of dwarf galaxies. With $R_c > 1\text{kpc}$ \cite{Zhang2018,Craciun2020,Dawoodbhoy2021}, then $\omega_0 \gtrsim  10^{-15}$, which is excluded at $2.4\sigma$ ($98.5\%$ confidence) by the large-scale observables considered in this work. The largest hydrostatic core radii allowed at $1\sigma$ is $R\approx 0.5\text{kpc}$, which is too small to account for the observed flat inner regions of low mass halos with SIBEC-DM near hydrostatic equilibrium. Some additional property of SIBEC-DM halos not captured by hydrostatic equilibrium appears to be needed to enlarge the halo cores in order to explain the cusp-core problem. However, ref. \cite{Shapiro2021} recently used perturbation theory and the Press-Schechter formalism \cite{Press1974} to compute the halo mass function of SIBEC-DM, and found, using constraints on the cut-off scale of structure formation in FDM as a proxy, that SIBEC-DM halo cores might be limited to $R_c$ as low as $10\text{pc}$, which poses a serious challenge for SIBEC-DM to explain the core-cusp problem, though the suppression in SIBEC-DM is softer compared to FDM, making a direct translation of constraints on FDM to SIBEC-DM difficult. FDM has a comoving Jeans' wavenumber that increases like $a^{1/4}$, much more slowly than SIBEC-DM which increases like $a$. SIBEC-DM perturbations of a given wavelength therefore become larger than the Jeans' length faster compared to FDM, and scales that were suppressed and Jeans filtered early can still recover and grow \cite{Shapiro2021}, leaving more low-mass halos in the late universe. Further efforts to test SIBEC-DM more directly against observables such as the halo mass function
likely requires fully 3D cosmological simulations of structure formation, but is expected to provide much more stringent constraints on SIBEC-DM compared to the large-scale observables used in this work, as shown by the work in ref. \cite{Shapiro2021}.

\begin{figure}
    \centering 
    \includegraphics[width=0.49\linewidth]{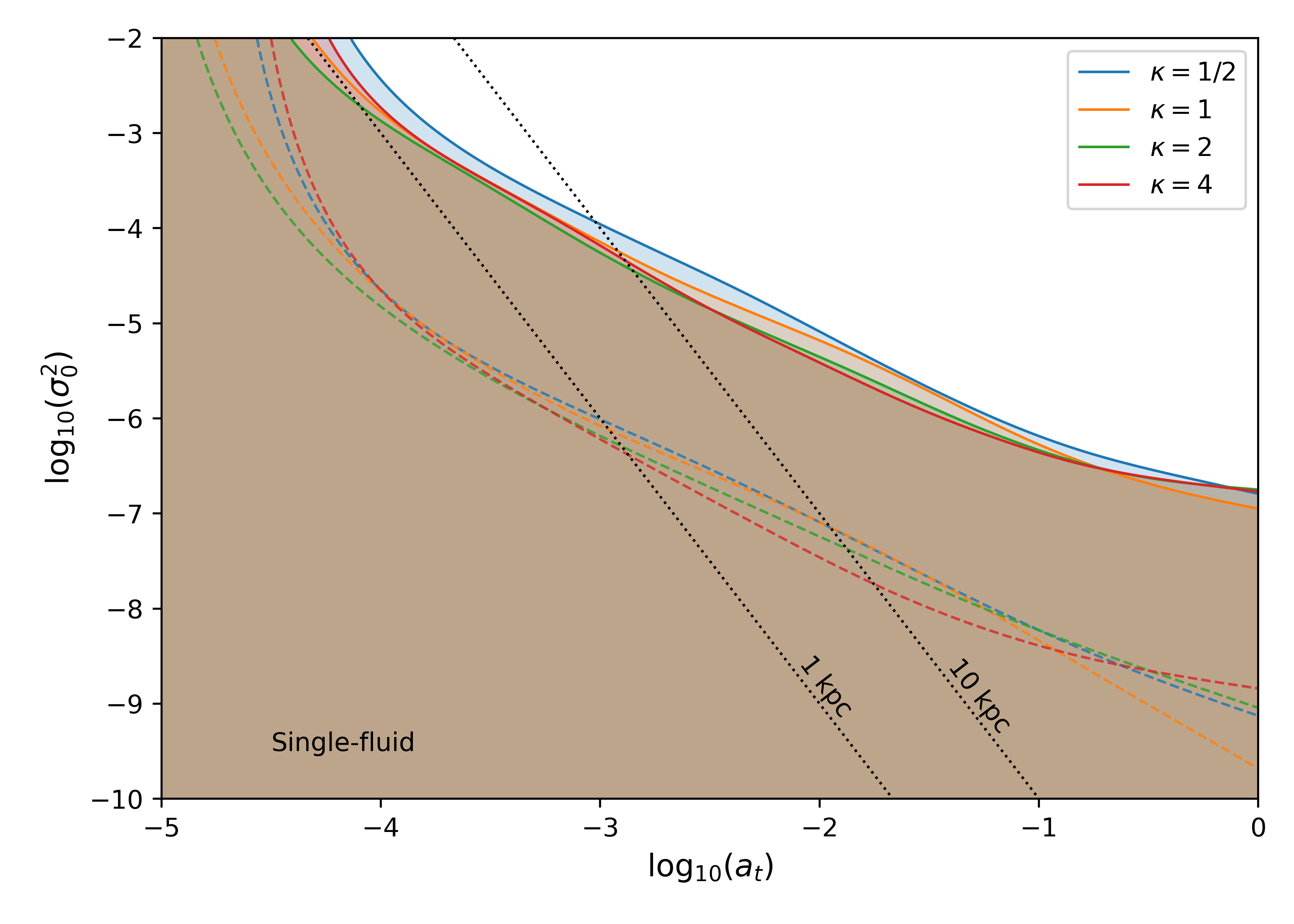}
    \includegraphics[width=0.49\linewidth]{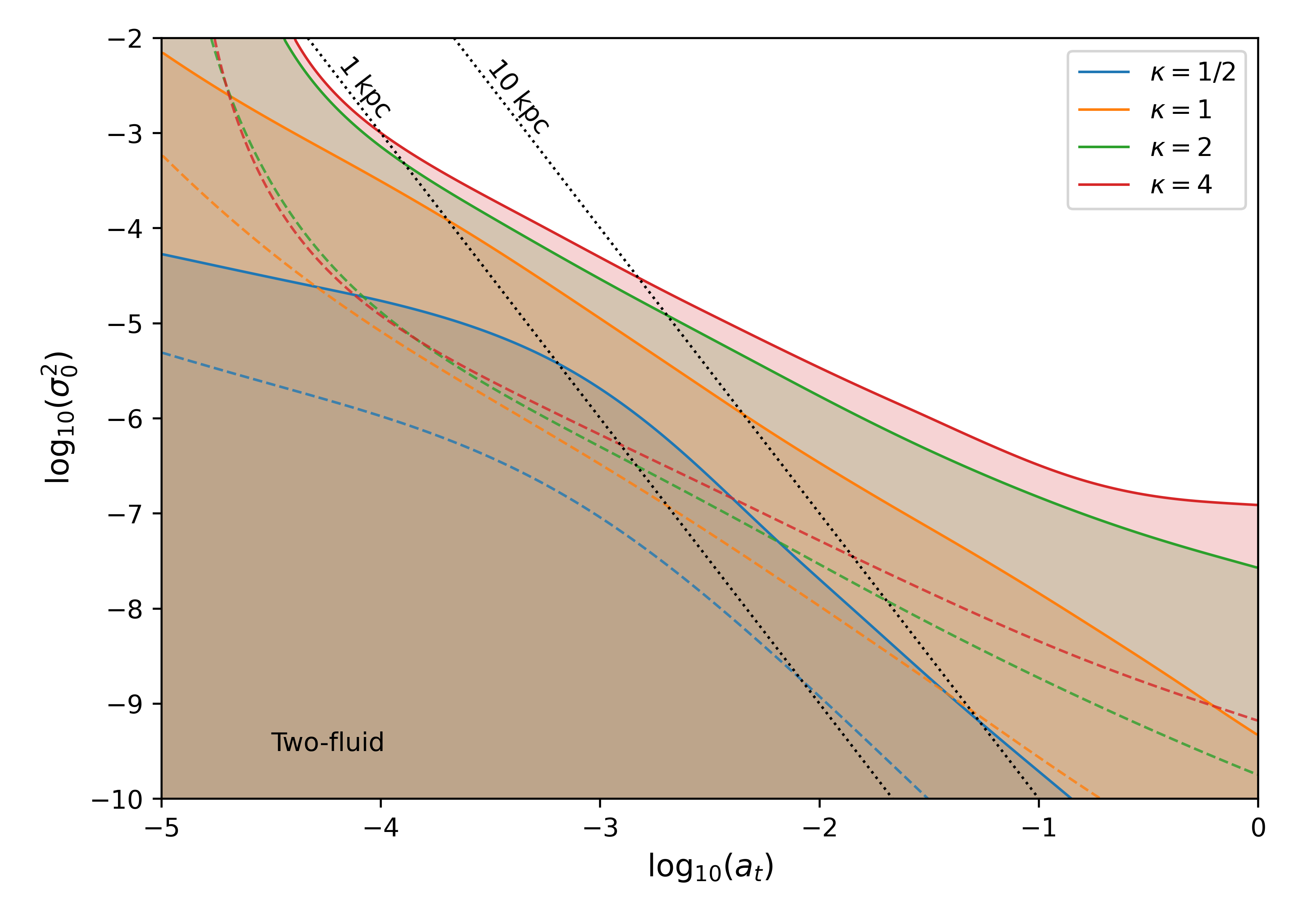}

    \caption{The 95\% (\textit{shaded contours}) and 68\% (\textit{colored dashed lines}) credible regions for $\sigma_0$-SIBEC in the single-fluid (\textit{left}), and the two-fluid (\textit{right}) case, using all the cosmological datasets (TTTEEE+lensing+BOSS+Pantheon). The parameters that yield hydrostatic core radii of $1$kpc and $10$kpc are shown in black dotted lines.}
\label{fig:mcmc_various_kappa_alldata_constant_w_SIBEC}
\end{figure}

The upper bound on $\omega_0$ can be relaxed considerably by having SIBEC-DM be the end product of some transition from an earlier phase, yet some general observations can be made. For instance, if SIBEC-DM is to provide a solution for the cusp-core problem in the late universe, with core radii larger than $1\text{kpc}$, then the transition time $a_t$ into the SIBEC-DM phase should not be much earlier than matter-radiation equality. Even transitions before $z\sim 600-700$ are slightly disfavored in this case, and requires a quite cold initial phase, with either $\sigma^2_0 \lesssim 10^{-6}$ for the initial constant equation of state, or $b^2 \lesssim 10^{-12}$ for WDM. We also find that $\sigma_0$-SIBEC-DM in the single-fluid approach does not depend much on the rate of the transition, while it does for cold-SIBEC-DM. This is due to the fact that changing the width of the crossover between two phases with $\omega\neq 0$ does not really change the overall shape of $\omega_{\chi}$ much, hence the effect on observables is small. Lowering the rate at which $\omega_{\chi}$ goes from zero to $\omega_{\text{SIBEC}}$, on the other hand, does have a large effect on observables, since lowering $\kappa$ in eq. \eqref{eq:cold_SIBEC_eos} increases $\omega_{\chi}$ significantly for $a<a_t$.

\begin{figure}
    \centering 
    \includegraphics[width=0.49\linewidth]{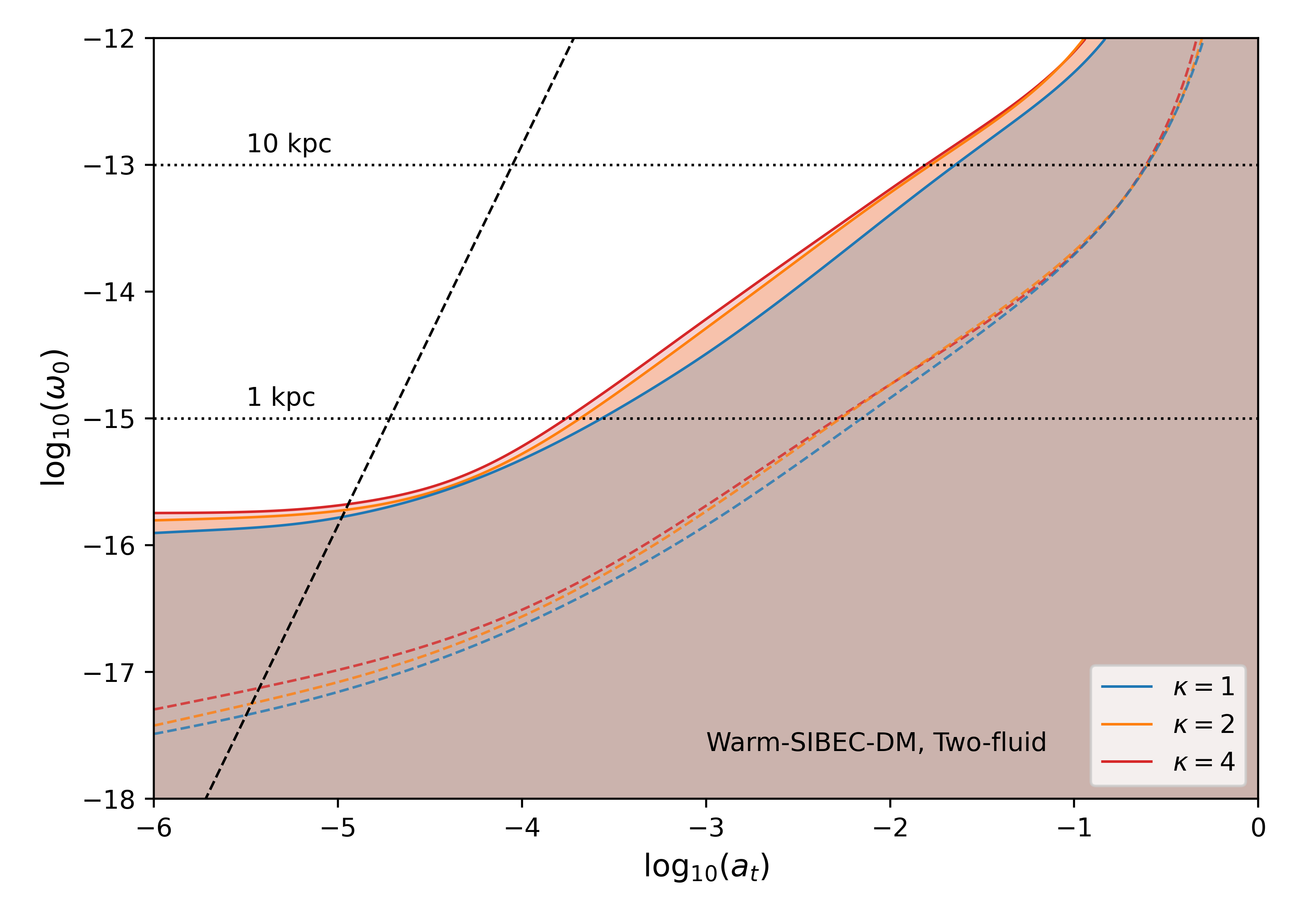}
    \includegraphics[width=0.49\linewidth]{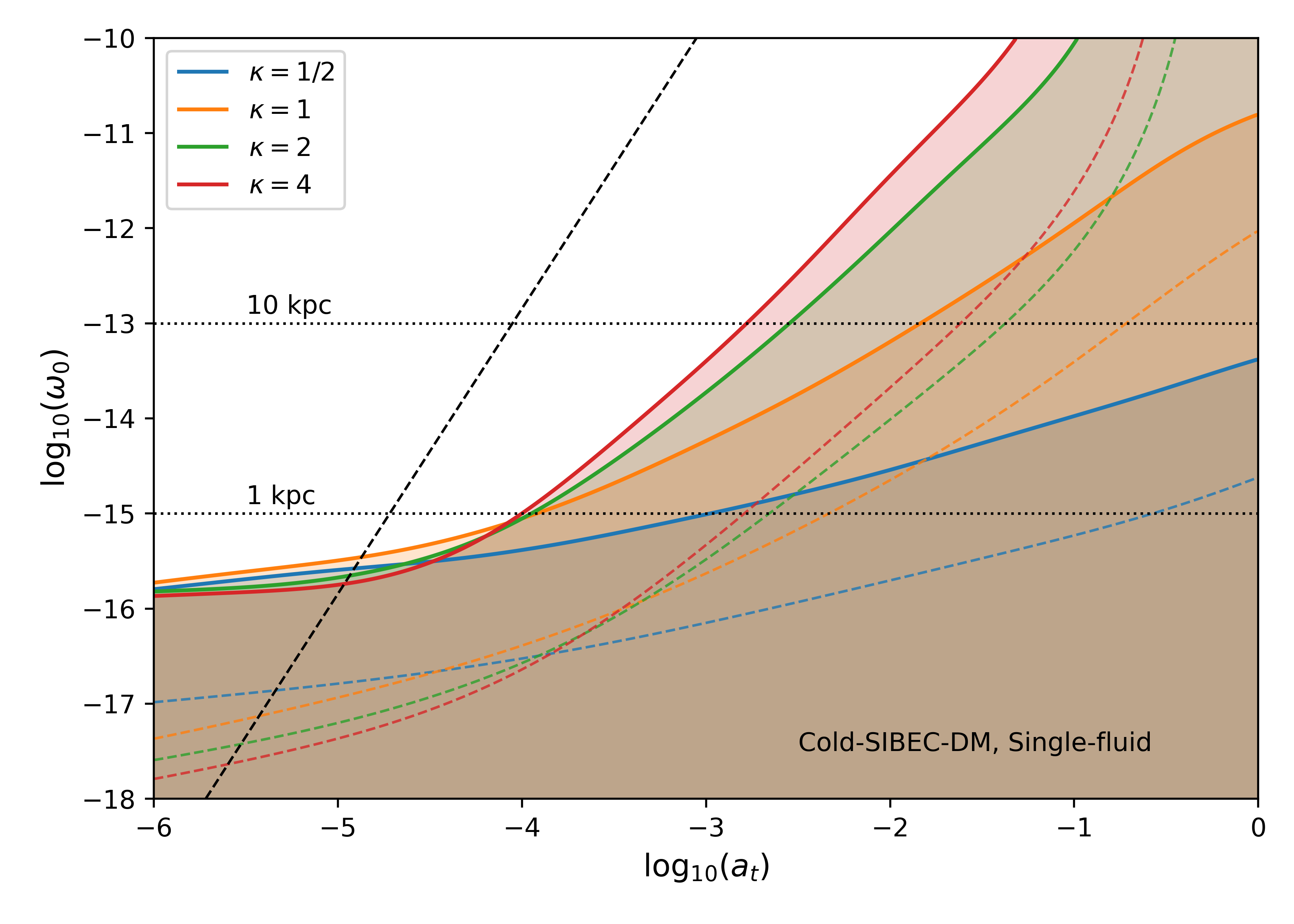}
    \includegraphics[width=0.49\linewidth]{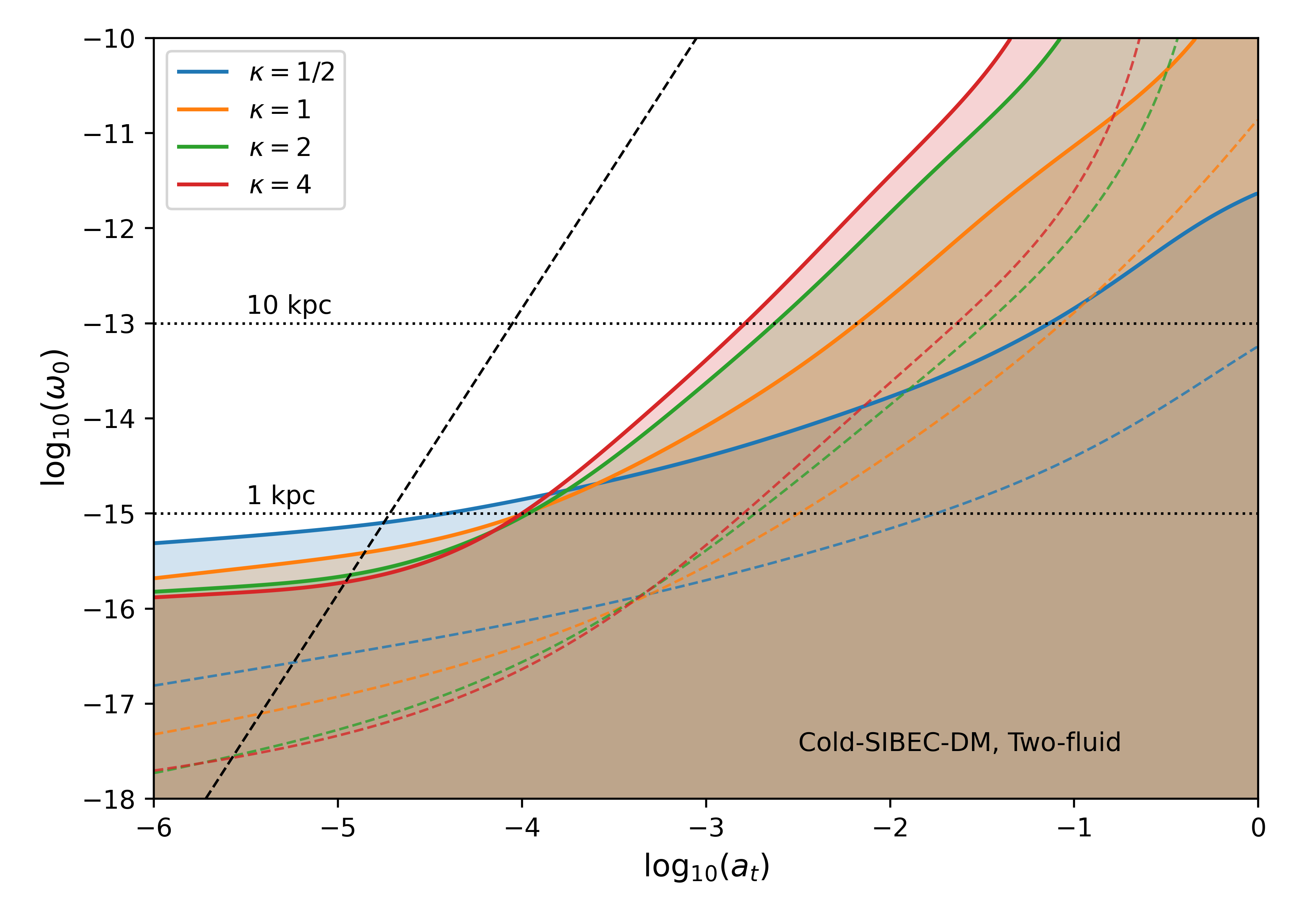}

    \caption{The 95\% (\textit{shaded contours}) and 68\% (\textit{colored dashed lines}) credible regions for two-fluid warm-SIBEC-DM (\textit{upper}) , single-fluid cold-SIBEC-DM (\textit{lower left}), and two-fluid cold-SIBEC-DM (\textit{lower right}), using all the cosmological datasets (TTTEEE+lensing+BOSS+Pantheon). The equation of state of SIBEC-DM today, $\omega_0$, for a DM halo at hydrostatic equilibrium with core size $1$kpc and $10$kpc (\textit{horizontal dotted lines}). The scale factor $a_{\text{nr}}$ where $\omega_{\text{SIBEC}} = 0.1$, i.e. starts to become non-relativistic, is shown with the dashed black line using the same axis as $\log_{10}(a_t)$. Transitions to the left of this line happen when the SIBEC phase is still radiation-like.}
\label{fig:mcmc_warm_and_cold_SIBEC}
\end{figure}

\section{Conclusion}
\label{sec:conclusions}
In this work we looked at the impact of scalar field DM with self-interactions---which we call SIBEC-DM---on large-scale observables, and placed bounds on the extra parameters introduced. We considered pure SIBEC-DM, which is fully described by a scalar-field Lagrangian with quartic self-interactions (though we did not specify what kind of non-thermal mechanism that produced the scalar field in the early universe, assuming only adiabatic initial conditions during radiation-domination). We also considered several scenarios in which SIBEC-DM is the end product of some transition from an initial normal fluid phase, which has previously been studied at the linear level using a Newtonian approach, in which the fluid pressure contributes to the dynamical evolution of the background. The transition into the SIBEC-fluid phase has also usually been approximated as nearly instantaneous \cite{Harko2011,Velteen2012,Freitas2013,Freitas2015}.
However, these approaches break down at scales close to and beyond the horizon, and fail to 
describe the evolution of structure from their primordial perturbations---seeded by inflation---to what we see today, which is necessary for accurately computing large-scale observables such as the CMB. In this work we therefore employed cosmological perturbation theory to compute these large-scale observables, and also allowed for the transition to take place over a long time. The resulting linear equations, as well as the changes to the background evolution, were implemented into the Boltzmann code CLASS, and the MCMC code MontePython used to derive constraints on SIBEC-DM.

The standard cosmological parameters were found to be unchanged from $\Lambda$CDM, and the CDM limit of the extra SIBEC-DM parameters to be favored by the linear probes, i.e. $\omega_0, \sigma_0, b \rightarrow 0$ and $a_t\rightarrow \infty$. In the simplest scenario with purely SIBEC-DM, we found an upper bound on the self-interaction of the scalar field of $\omega_0 < 3.0\times 10^{-16}$ at $1\sigma$, and that the self-interaction necessary for solving the small-scale discrepancies in $\Lambda$CDM, with $R_c\gtrsim 1\text{kpc}$, is disfavored by the present work at $2.4\sigma$, or $98.5\%$ confidence. Throughout this work, however, the initial stiff phase of self-interacting scalar field DM \cite{Li2014} was neglected, which would give different initial conditions since the universe would be dominated by the stiff DM field instead of radiation at very early times. The consequences of this on the present results would be interesting to investigate in the future.

The above constraint is relaxed if SIBEC-DM is the end product of some transition from an earlier phase that is sufficiently cold, and has been the working assumption of several studies \cite{Harko2011,Harko2012,Velteen2012,Freitas2013,Bettoni2014,Freitas2015}. We derived bounds on the phenomenology of such a transition for multiple scenarios, with DM starting in out in a phase that is either CDM-like, has a constant equation of state, or that is WDM-like. A general feature is that for SIBEC-DM to be a solution to e.g. the core-cusp problem, with halo core radii from hydrostatic equilibrium larger than around $1\text{kpc}$, then transitions shortly before matter-radiation equality and earlier are disfavored at around $2\sigma$ in all the cases considered, and at $1\sigma$ the lower bound on $a_t$ is pushed to after recombination. Transitions at later times result in SIBEC-DM that is indistinguishable from CDM in the large-scale observables considered in work. 

Any successful DM model must have a description for the evolution of DM throughout cosmic history that is consistent with observables on all scales, large and small, thus investigating the properties of SIBEC-DM in the late universe in the non-linear regime is also of great interest, particularly going beyond hydrostatic considerations and simulating the formation and evolution of structure in a SIBEC-DM universe. The first step in this direction was recently made in refs. \cite{Dawoodbhoy2021,Shapiro2021}, in which the collapse of spherically symmetric SIBEC-DM overdensities was studied. In a non-cosmological setup, self-interactions that results in $R_c \gtrsim 1\text{kpc}$ was again found to be needed to solve the small-scale issues of $\Lambda$CDM. In a cosmological setup, on the other hand, such a large self-interaction was found to overly suppress the formation of low-mass halos via early Jeans filtering of high frequency $k$-modes, drawing a similar conclusion as us that $R_c\gtrsim 1\text{kpc}$ is in conflict with observations. In fact, they find that $R_c$ should be smaller than $10\text{pc}$ in order to not overly suppressing the abundance of low-mass halos, far below the value generally though to be needed for the cusp-core problem. The dynamical evolution of SIBEC-DM structures, as well as growth through mergers, is not captured under the assumption of spherical symmetry, or linear perturbation theory, so fully 3D large-scale cosmological simulations is therefore the natural next step for investigating and fully utilizing late-time probes to constrain SIBEC-DM, as well as to obtain more realistic profiles and core radii for SIBEC-DM halos.

\acknowledgments{We thank the Research Council of Norway for their support. We are also grateful for the feedback from the anonymous referee, which helped improve the manuscript considerably. Computations were performed on resources provided by UNINETT Sigma2 -- the National Infrastructure for High Performance Computing and Data Storage in Norway.}


\begin{thebibliography}{98}
\expandafter\ifx\csname natexlab\endcsname\relax\def\natexlab#1{#1}\fi

\bibitem[{Ade {et~al.}(2016)Ade, Aghanim, Arnaud, Ashdown, Aumont, Baccigalupi,
  Banday, Barreiro, Bartlett, Bartolo, Battaner, Battye, Benabed, Benoît,
  Benoit-Lévy, Bernard, Bersanelli, Bielewicz, Bock, Bonaldi, Bonavera, Bond,
  Borrill, Bouchet, Boulanger, Bucher, Burigana, Butler, Calabrese, Cardoso,
  Catalano, Challinor, Chamballu, Chary, Chiang, Chluba, Christensen, Church,
  Clements, Colombi, Colombo, Combet, Coulais, Crill, Curto, Cuttaia, Danese,
  Davies, Davis, Bernardis, Rosa, Zotti, Delabrouille, Désert, Valentino,
  Dickinson, Diego, Dolag, Dole, Donzelli, Doré, Douspis, Ducout, Dunkley,
  Dupac, Efstathiou, Elsner, Enßlin, Eriksen, Farhang, Fergusson, Finelli,
  Forni, Frailis, Fraisse, Franceschi, Frejsel, Galeotta, Galli, Ganga,
  Gauthier, Gerbino, Ghosh, Giard, Giraud-Héraud, Giusarma, Gjerløw,
  González-Nuevo, Górski, Gratton, Gregorio, Gruppuso, Gudmundsson, Hamann,
  Hansen, Hanson, Harrison, Helou, Henrot-Versillé, Hernández-Monteagudo,
  Herranz, Hildebrandt, Hivon, Hobson, Holmes, Hornstrup, Hovest, Huang,
  Huffenberger, Hurier, Jaffe, Jaffe, Jones, Juvela, Keihänen, Keskitalo,
  Kisner, Kneissl, Knoche, Knox, Kunz, Kurki-Suonio, Lagache, Lähteenmäki,
  Lamarre, Lasenby, Lattanzi, Lawrence, Leahy, Leonardi, Lesgourgues, Levrier,
  Lewis, Liguori, Lilje, Linden-Vørnle, López-Caniego, Lubin, Macías-Pérez,
  Maggio, Maino, Mandolesi, Mangilli, Marchini, Maris, Martin, Martinelli,
  Martínez-González, Masi, Matarrese, McGehee, Meinhold, Melchiorri, Melin,
  Mendes, Mennella, Migliaccio, Millea, Mitra, Miville-Deschênes, Moneti,
  Montier, Morgante, Mortlock, Moss, Munshi, Murphy, Naselsky, Nati, Natoli,
  Netterfield, Nørgaard-Nielsen, Noviello, Novikov, Novikov, Oxborrow, Paci,
  Pagano, Pajot, Paladini, Paoletti, Partridge, Pasian, Patanchon, Pearson,
  Perdereau, Perotto, Perrotta, Pettorino, Piacentini, Piat, Pierpaoli,
  Pietrobon, Plaszczynski, Pointecouteau, Polenta, Popa, Pratt, Prézeau,
  Prunet, Puget, Rachen, Reach, Rebolo, Reinecke, Remazeilles, Renault, Renzi,
  Ristorcelli, Rocha, Rosset, Rossetti, Roudier, d’Orfeuil, Rowan-Robinson,
  Rubiño-Martín, Rusholme, Said, Salvatelli, Salvati, Sandri, Santos,
  Savelainen, Savini, Scott, Seiffert, Serra, Shellard, Spencer, Spinelli,
  Stolyarov, Stompor, Sudiwala, Sunyaev, Sutton, Suur-Uski, Sygnet, Tauber,
  Terenzi, Toffolatti, Tomasi, Tristram, Trombetti, Tucci, Tuovinen, Türler,
  Umana, Valenziano, Valiviita, Tent, Vielva, Villa, Wade, Wandelt, Wehus,
  White, White, Wilkinson, Yvon, Zacchei, \& Zonca}]{Planck2015}
Ade, P. A. R., Aghanim, N., Arnaud, M., {et~al.} 2016, Astronomy \&
  Astrophysics, 594, A13, publisher: EDP Sciences

\bibitem[{Aghanim {et~al.}(2020)Aghanim, Akrami, Ashdown, Aumont, Baccigalupi,
  Ballardini, Banday, Barreiro, Bartolo, Basak, Battye, Benabed, Bernard,
  Bersanelli, Bielewicz, Bock, Bond, Borrill, Bouchet, Boulanger, Bucher,
  Burigana, Butler, Calabrese, Cardoso, Carron, Challinor, Chiang, Chluba,
  Colombo, Combet, Contreras, Crill, Cuttaia, Bernardis, Zotti, Delabrouille,
  Delouis, Valentino, Diego, Doré, Douspis, Ducout, Dupac, Dusini, Efstathiou,
  Elsner, Enßlin, Eriksen, Fantaye, Farhang, Fergusson, Fernandez-Cobos,
  Finelli, Forastieri, Frailis, Fraisse, Franceschi, Frolov, Galeotta, Galli,
  Ganga, Génova-Santos, Gerbino, Ghosh, González-Nuevo, Górski, Gratton,
  Gruppuso, Gudmundsson, Hamann, Handley, Hansen, Herranz, Hildebrandt, Hivon,
  Huang, Jaffe, Jones, Karakci, Keihänen, Keskitalo, Kiiveri, Kim, Kisner,
  Knox, Krachmalnicoff, Kunz, Kurki-Suonio, Lagache, Lamarre, Lasenby,
  Lattanzi, Lawrence, Jeune, Lemos, Lesgourgues, Levrier, Lewis, Liguori,
  Lilje, Lilley, Lindholm, López-Caniego, Lubin, Ma, Macías-Pérez, Maggio,
  Maino, Mandolesi, Mangilli, Marcos-Caballero, Maris, Martin, Martinelli,
  Martínez-González, Matarrese, Mauri, McEwen, Meinhold, Melchiorri,
  Mennella, Migliaccio, Millea, Mitra, Miville-Deschênes, Molinari, Montier,
  Morgante, Moss, Natoli, Nørgaard-Nielsen, Pagano, Paoletti, Partridge,
  Patanchon, Peiris, Perrotta, Pettorino, Piacentini, Polastri, Polenta, Puget,
  Rachen, Reinecke, Remazeilles, Renzi, Rocha, Rosset, Roudier,
  Rubiño-Martín, Ruiz-Granados, Salvati, Sandri, Savelainen, Scott, Shellard,
  Sirignano, Sirri, Spencer, Sunyaev, Suur-Uski, Tauber, Tavagnacco, Tenti,
  Toffolatti, Tomasi, Trombetti, Valenziano, Valiviita, Tent, Vibert, Vielva,
  Villa, Vittorio, Wandelt, Wehus, White, White, Zacchei, \&
  Zonca}]{Planck2020}
Aghanim, N., Akrami, Y., Ashdown, M., {et~al.} 2020, Astronomy \& Astrophysics,
  641, A6, publisher: EDP Sciences

\bibitem[{Alam {et~al.}(2017)Alam, Ata, Bailey, Beutler, Bizyaev, Blazek,
  Bolton, Brownstein, Burden, Chuang, Comparat, Cuesta, Dawson, Eisenstein,
  Escoffier, Gil-Marín, Grieb, Hand, Ho, Kinemuchi, Kirkby, Kitaura,
  Malanushenko, Malanushenko, Maraston, McBride, Nichol, Olmstead, Oravetz,
  Padmanabhan, Palanque-Delabrouille, Pan, Pellejero-Ibanez, Percival,
  Petitjean, Prada, Price-Whelan, Reid, Rodríguez-Torres, Roe, Ross, Ross,
  Rossi, Rubiño-Martín, Saito, Salazar-Albornoz, Samushia, Sánchez,
  Satpathy, Schlegel, Schneider, Scóccola, Seo, Sheldon, Simmons, Slosar,
  Strauss, Swanson, Thomas, Tinker, Tojeiro, Magaña, Vazquez, Verde, Wake,
  Wang, Weinberg, White, Wood-Vasey, Yèche, Zehavi, Zhai, \& Zhao}]{Alam2017}
Alam, S., Ata, M., Bailey, S., {et~al.} 2017, Monthly Notices of the Royal
  Astronomical Society, 470, 2617

\bibitem[{Arbey {et~al.}(2002)Arbey, Lesgourgues, \& Salati}]{Arbey2002}
Arbey, A., Lesgourgues, J., \& Salati, P. 2002, Physical Review D, 65, 083514,
  publisher: American Physical Society

\bibitem[{Arbey {et~al.}(2003)Arbey, Lesgourgues, \& Salati}]{Arbey2003}
Arbey, A., Lesgourgues, J., \& Salati, P. 2003, Physical Review D, 68, 023511,
  publisher: American Physical Society

\bibitem[{Armengaud {et~al.}(2017)Armengaud, Palanque-Delabrouille, Yèche,
  Marsh, \& Baur}]{Armengaud2017}
Armengaud, E., Palanque-Delabrouille, N., Yèche, C., Marsh, D. J.~E., \& Baur,
  J. 2017, Monthly Notices of the Royal Astronomical Society, 471, 4606

\bibitem[{Audren {et~al.}(2013)Audren, Lesgourgues, Benabed, \&
  Prunet}]{Audren2012}
Audren, B., Lesgourgues, J., Benabed, K., \& Prunet, S. 2013, JCAP, 1302, 001

\bibitem[{Berezhiani \& Khoury(2015)}]{Berezhiani2015}
Berezhiani, L. \& Khoury, J. 2015, Phys. Rev. D, 92, 103510

\bibitem[{Bettoni {et~al.}(2014)Bettoni, Colombo, \& Liberati}]{Bettoni2014}
Bettoni, D., Colombo, M., \& Liberati, S. 2014, Journal of Cosmology and
  Astroparticle Physics, 2014, 004, publisher: IOP Publishing

\bibitem[{Billard {et~al.}(2014)Billard, Figueroa-Feliciano, \&
  Strigari}]{Billard2014}
Billard, J., Figueroa-Feliciano, E., \& Strigari, L. 2014, Physical Review D,
  89, 023524, publisher: American Physical Society

\bibitem[{Blas {et~al.}(2011)Blas, Lesgourgues, \& Tram}]{Blas2011}
Blas, D., Lesgourgues, J., \& Tram, T. 2011, Journal of Cosmology and
  Astroparticle Physics, 2011, 034, publisher: IOP Publishing

\bibitem[{Boyarsky {et~al.}(2019)Boyarsky, Drewes, Lasserre, Mertens, \&
  Ruchayskiy}]{Boyarsky2019}
Boyarsky, A., Drewes, M., Lasserre, T., Mertens, S., \& Ruchayskiy, O. 2019,
  Progress in Particle and Nuclear Physics, 104, 1

\bibitem[{Boylan-Kolchin {et~al.}(2011)Boylan-Kolchin, Bullock, \&
  Kaplinghat}]{Boylan-Kolchin2011}
Boylan-Kolchin, M., Bullock, J.~S., \& Kaplinghat, M. 2011, Monthly Notices of
  the Royal Astronomical Society: Letters, 415, L40

\bibitem[{Brinckmann \& Lesgourgues(2018)}]{Brinckmann2018}
Brinckmann, T. \& Lesgourgues, J. 2018, MontePython 3: boosted MCMC sampler and
  other features

\bibitem[{Bringmann {et~al.}(2018)Bringmann, Kahlhoefer, Schmidt-Hoberg, \&
  Walia}]{Bringmann2018}
Bringmann, T., Kahlhoefer, F., Schmidt-Hoberg, K., \& Walia, P. 2018, Physical
  Review D, 98, 023543, publisher: American Physical Society

\bibitem[{Bullock \& Boylan-Kolchin(2017)}]{Bullock2017}
Bullock, J.~S. \& Boylan-Kolchin, M. 2017, Annual Review of Astronomy and
  Astrophysics, 55, 343

\bibitem[{Böhmer \& Harko(2007)}]{Bohmer2007}
Böhmer, C.~G. \& Harko, T. 2007, Journal of Cosmology and Astroparticle
  Physics, 2007, 025, publisher: IOP Publishing

\bibitem[{Chavanis(2011)}]{Chavanis2011}
Chavanis, P.-H. 2011, Physical Review D, 84, 043531, publisher: American
  Physical Society

\bibitem[{Chavanis \& Delfini(2011)}]{Chavanis2011b}
Chavanis, P.-H. \& Delfini, L. 2011, Physical Review D, 84, 043532, publisher:
  American Physical Society

\bibitem[{Crăciun \& Harko(2020)}]{Craciun2020}
Crăciun, M. \& Harko, T. 2020, arXiv:2007.12222 [astro-ph, physics:gr-qc,
  physics:hep-th], arXiv: 2007.12222

\bibitem[{Davis {et~al.}(1985)Davis, Efstathiou, Frenk, \& White}]{Davis1985}
Davis, M., Efstathiou, G., Frenk, C.~S., \& White, S. D.~M. 1985, The
  Astrophysical Journal, 292, 371

\bibitem[{Dawoodbhoy {et~al.}(2021)Dawoodbhoy, Shapiro, \&
  Rindler-Daller}]{Dawoodbhoy2021}
Dawoodbhoy, T., Shapiro, P.~R., \& Rindler-Daller, T. 2021, Monthly Notices of
  the Royal Astronomical Society, 506, 2418

\bibitem[{De~Berredo-Peixoto {et~al.}(2005)De~Berredo-Peixoto, Shapiro, \&
  Sobreira}]{Berredo-Peixoto2005}
De~Berredo-Peixoto, G., Shapiro, I.~L., \& Sobreira, F. 2005, Modern Physics
  Letters A, 20, 2723, publisher: World Scientific Publishing Co.

\bibitem[{de~Freitas \& Velten(2015)}]{Freitas2015}
de~Freitas, R.~C. \& Velten, H. 2015, The European Physical Journal C, 75, 597

\bibitem[{Del~Popolo \& Le~Delliou(2017)}]{DelPopolo2017}
Del~Popolo, A. \& Le~Delliou, M. 2017, Galaxies, 5

\bibitem[{Dodelson(2003)}]{Dodelson2003}
Dodelson, S. 2003, {Modern Cosmology} (Amsterdam: Academic Press)

\bibitem[{dos Reis \& Shapiro(2018)}]{Reis2018}
dos Reis, S.~C. \& Shapiro, I.~L. 2018, The European Physical Journal C, 78,
  145

\bibitem[{Erken {et~al.}(2012)Erken, Sikivie, Tam, \& Yang}]{Erken2012}
Erken, O., Sikivie, P., Tam, H., \& Yang, Q. 2012, Physical Review D, 85,
  063520, publisher: American Physical Society

\bibitem[{Fabris {et~al.}(2009)Fabris, Shapiro, \& Sobreira}]{Fabris2009}
Fabris, J.~C., Shapiro, I.~L., \& Sobreira, F. 2009, 2009, 001, publisher: IOP
  Publishing

\bibitem[{Fabris {et~al.}(2012)Fabris, Shapiro, \&
  Velasquez-Toribio}]{Fabris2012}
Fabris, J.~C., Shapiro, I.~L., \& Velasquez-Toribio, A.~M. 2012, Physical
  Review D, 85, 023506, publisher: American Physical Society

\bibitem[{Faraoni(2012)}]{Faraoni2012}
Faraoni, V. 2012, Physical Review D, 85, 024040, publisher: American Physical
  Society

\bibitem[{Ferreira(2020)}]{Ferreira2020}
Ferreira, E. G.~M. 2020, arXiv:2005.03254 [astro-ph, physics:cond-mat,
  physics:gr-qc, physics:hep-th], arXiv: 2005.03254

\bibitem[{Ferreira {et~al.}(2019)Ferreira, Franzmann, Khoury, \&
  Brandenberger}]{Ferreira2019}
Ferreira, E. G.~M., Franzmann, G., Khoury, J., \& Brandenberger, R. 2019,
  Journal of Cosmology and Astroparticle Physics, 2019, 027, arXiv: 1810.09474

\bibitem[{Freitas \& Gonçalves(2013)}]{Freitas2013}
Freitas, R.~C. \& Gonçalves, S. V.~B. 2013, Journal of Cosmology and
  Astroparticle Physics, 2013, 049, publisher: IOP Publishing

\bibitem[{Garrison-Kimmel {et~al.}(2014)Garrison-Kimmel, Boylan-Kolchin,
  Bullock, \& Lee}]{Garrison-Kimmel2014}
Garrison-Kimmel, S., Boylan-Kolchin, M., Bullock, J.~S., \& Lee, K. 2014,
  Monthly Notices of the Royal Astronomical Society, 438, 2578

\bibitem[{Gelman \& Rubin(1992)}]{Gelman1992}
Gelman, A. \& Rubin, D.~B. 1992, Statistical Science, 7, 457

\bibitem[{Goodman(2000)}]{Goodman2000}
Goodman, J. 2000, New Astronomy, 5, 103

\bibitem[{Guth {et~al.}(2015)Guth, Hertzberg, \& Prescod-Weinstein}]{Guth2015}
Guth, A.~H., Hertzberg, M.~P., \& Prescod-Weinstein, C. 2015, Physical Review
  D, 92, 103513, publisher: American Physical Society

\bibitem[{Harko(2011)}]{Harko2011}
Harko, T. 2011, Physical Review D, 83, 123515

\bibitem[{Harko \& Mocanu(2012)}]{Harko2012}
Harko, T. \& Mocanu, G. 2012, Physical Review D, 85, 084012

\bibitem[{Harvey {et~al.}(2015)Harvey, Massey, Kitching, Taylor, \&
  Tittley}]{Harvey2015}
Harvey, D., Massey, R., Kitching, T., Taylor, A., \& Tittley, E. 2015, Science,
  347, 1462

\bibitem[{Hipólito-Ricaldi {et~al.}(2018)Hipólito-Ricaldi, vom Marttens,
  Fabris, Shapiro, \& Casarini}]{Hipolito-Ricaldi2018}
Hipólito-Ricaldi, W.~S., vom Marttens, R.~F., Fabris, J.~C., Shapiro, I.~L.,
  \& Casarini, L. 2018, The European Physical Journal C, 78, 365

\bibitem[{Hlozek {et~al.}(2015)Hlozek, Grin, Marsh, \& Ferreira}]{Hlozek2015}
Hlozek, R., Grin, D., Marsh, D.~J., \& Ferreira, P.~G. 2015, Physical Review D,
  91, 103512, publisher: American Physical Society

\bibitem[{Hložek {et~al.}(2018)Hložek, Marsh, \& Grin}]{Hlozek2018}
Hložek, R., Marsh, D. J.~E., \& Grin, D. 2018, Monthly Notices of the Royal
  Astronomical Society, 476, 3063

\bibitem[{Hu(1998)}]{Hu1998}
Hu, W. 1998, The Astrophysical Journal, 506, 485, publisher: IOP Publishing

\bibitem[{Hwang \& Noh(2009)}]{Hwang2009}
Hwang, J.-c. \& Noh, H. 2009, Physics Letters B, 680, 1

\bibitem[{Ilić {et~al.}(2020)Ilić, Kopp, Skordis, \& Thomas}]{Ilic2020}
Ilić, S., Kopp, M., Skordis, C., \& Thomas, D.~B. 2020, arXiv:2004.09572
  [astro-ph], arXiv: 2004.09572

\bibitem[{Iršič {et~al.}(2017)Iršič, Viel, Haehnelt, Bolton, \&
  Becker}]{Irsic2017}
Iršič, V., Viel, M., Haehnelt, M.~G., Bolton, J.~S., \& Becker, G.~D. 2017,
  Physical Review Letters, 119, 031302, publisher: American Physical Society

\bibitem[{Joudaki {et~al.}(2020)Joudaki, Ferreira, Lima, \&
  Winther}]{Joudaki2020}
Joudaki, S., Ferreira, P.~G., Lima, N.~A., \& Winther, H.~A. 2020,
  arXiv:2010.15278 [astro-ph, physics:gr-qc], arXiv: 2010.15278

\bibitem[{Khoury(2016)}]{Khoury2016}
Khoury, J. 2016, Phys. Rev. D, 93, 103533

\bibitem[{Kobayashi {et~al.}(2017)Kobayashi, Murgia, De~Simone, Iršič, \&
  Viel}]{Kobayashi2017}
Kobayashi, T., Murgia, R., De~Simone, A., Iršič, V., \& Viel, M. 2017,
  Physical Review D, 96, 123514, publisher: American Physical Society

\bibitem[{Kopp {et~al.}(2016)Kopp, Skordis, \& Thomas}]{Kopp2016}
Kopp, M., Skordis, C., \& Thomas, D.~B. 2016, Physical Review D, 94, 043512,
  arXiv: 1605.00649

\bibitem[{Laguë {et~al.}(2021)Laguë, Bond, Hložek, Rogers, Marsh, \&
  Grin}]{Lague2021}
Laguë, A., Bond, J.~R., Hložek, R., {et~al.} 2021, arXiv:2104.07802
  [astro-ph], arXiv: 2104.07802

\bibitem[{Lancaster {et~al.}(2020)Lancaster, Giovanetti, Mocz, Kahn, Lisanti,
  \& Spergel}]{Lancaster2020}
Lancaster, L., Giovanetti, C., Mocz, P., {et~al.} 2020, Journal of Cosmology
  and Astroparticle Physics, 2020, 001, publisher: IOP Publishing

\bibitem[{Lelli {et~al.}(2016)Lelli, McGaugh, \& Schombert}]{Lelli2016}
Lelli, F., McGaugh, S.~S., \& Schombert, J.~M. 2016, The Astronomical Journal,
  152, 157, publisher: American Astronomical Society

\bibitem[{Li {et~al.}(2014)Li, Rindler-Daller, \& Shapiro}]{Li2014}
Li, B., Rindler-Daller, T., \& Shapiro, P.~R. 2014, Physical Review D, 89,
  083536, publisher: American Physical Society

\bibitem[{Ma \& Bertschinger(1995)}]{Ma1995}
Ma, C.-P. \& Bertschinger, E. 1995, The Astrophysical Journal, 455, 7

\bibitem[{Madelung(1926)}]{Madelung1926}
Madelung, E. 1926, Naturwissenschaften, 14, 1004

\bibitem[{Marsh(2016)}]{Marsh2016}
Marsh, D. J.~E. 2016, Physics Reports, 643, 1

\bibitem[{Matos \& Arturo Ureña-López(2001)}]{Matos2001}
Matos, T. \& Arturo Ureña-López, L. 2001, Physical Review D, 63, 063506,
  publisher: American Physical Society

\bibitem[{May \& Springel(2021)}]{May2021}
May, S. \& Springel, V. 2021, arXiv:2101.01828 [astro-ph, physics:gr-qc],
  arXiv: 2101.01828

\bibitem[{Medeiros(2012)}]{Medeiros2012}
Medeiros, L.~G. 2012, Modern Physics Letters A, 27, 1250194, publisher: World
  Scientific Publishing Co.

\bibitem[{Mina {et~al.}(2020{\natexlab{a}})Mina, Mota, \& Winther}]{Mina2020}
Mina, M., Mota, D.~F., \& Winther, H.~A. 2020{\natexlab{a}}, Astronomy \&
  Astrophysics, 641, A107, publisher: EDP Sciences

\bibitem[{Mina {et~al.}(2020{\natexlab{b}})Mina, Mota, \& Winther}]{Mina2020b}
Mina, M., Mota, D.~F., \& Winther, H.~A. 2020{\natexlab{b}}, arXiv:2007.04119
  [astro-ph, physics:gr-qc], arXiv: 2007.04119

\bibitem[{Mocz {et~al.}(2017)Mocz, Vogelsberger, Robles, Zavala,
  Boylan-Kolchin, Fialkov, \& Hernquist}]{Mocz2017}
Mocz, P., Vogelsberger, M., Robles, V.~H., {et~al.} 2017, Monthly Notices of
  the Royal Astronomical Society, 471, 4559

\bibitem[{Monroe \& Fisher(2007)}]{Monroe2007}
Monroe, J. \& Fisher, P. 2007, Physical Review D, 76, 033007, publisher:
  American Physical Society

\bibitem[{Nollett \& Steigman(2015)}]{Nollett2015}
Nollett, K.~M. \& Steigman, G. 2015, Physical Review D, 91, 083505, publisher:
  American Physical Society

\bibitem[{Nori \& Baldi(2018)}]{Nori2018}
Nori, M. \& Baldi, M. 2018, Monthly Notices of the Royal Astronomical Society,
  478, 3935, publisher: Oxford Academic

\bibitem[{Nori \& Baldi(2020)}]{Nori2020}
Nori, M. \& Baldi, M. 2020, arXiv:2007.01316 [astro-ph], arXiv: 2007.01316

\bibitem[{O’Hare(2020)}]{Ohare2020}
O’Hare, C.~A. 2020, Physical Review D, 102, 063024, publisher: American
  Physical Society

\bibitem[{Park {et~al.}(2012)Park, Hwang, \& Noh}]{Park2012}
Park, C.-G., Hwang, J.-c., \& Noh, H. 2012, Physical Review D, 86, 083535,
  publisher: American Physical Society

\bibitem[{Peebles(2000)}]{Peebles2000}
Peebles, P. J.~E. 2000, The Astrophysical Journal, 534, L127, publisher: IOP
  Publishing

\bibitem[{Percival {et~al.}(2001)Percival, Baugh, Bland-Hawthorn, Bridges,
  Cannon, Cole, Colless, Collins, Couch, Dalton, De~Propris, Driver,
  Efstathiou, Ellis, Frenk, Glazebrook, Jackson, Lahav, Lewis, Lumsden, Maddox,
  Moody, Norberg, Peacock, Peterson, Sutherland, \& Taylor}]{Percival2001}
Percival, W.~J., Baugh, C.~M., Bland-Hawthorn, J., {et~al.} 2001, Monthly
  Notices of the Royal Astronomical Society, 327, 1297, publisher: Oxford
  Academic

\bibitem[{Pitaevskii \& Stringari(2016)}]{Pitaevskii2016}
Pitaevskii, L.~P. \& Stringari, S. 2016, Bose-Einstein Condensation and
  Superfluidity (Great Clarendon Street, Oxford, United Kingdom: Oxford
  University Press)

\bibitem[{Pordeus-da Silva {et~al.}(2019)Pordeus-da Silva, Batista, \&
  Medeiros}]{Silva2019}
Pordeus-da Silva, G., Batista, R.~C., \& Medeiros, L.~G. 2019, 2019, 043,
  publisher: IOP Publishing

\bibitem[{{Press} \& {Schechter}(1974)}]{Press1974}
{Press}, W.~H. \& {Schechter}, P. 1974, The Astrophysical Journal, 187, 425

\bibitem[{Riess {et~al.}(2016)Riess, Macri, Hoffmann, Scolnic, Casertano,
  Filippenko, Tucker, Reid, Jones, Silverman, Chornock, Challis, Yuan, Brown,
  \& Foley}]{Riess2016}
Riess, A.~G., Macri, L.~M., Hoffmann, S.~L., {et~al.} 2016, The Astrophysical
  Journal, 826, 56, publisher: American Astronomical Society

\bibitem[{Rindler-Daller \& Shapiro(2012)}]{Rindler-Daller2012}
Rindler-Daller, T. \& Shapiro, P.~R. 2012, Monthly Notices of the Royal
  Astronomical Society, 422, 135

\bibitem[{Rindler-Daller \& Shapiro(2014)}]{Rindler-Daller2014}
Rindler-Daller, T. \& Shapiro, P.~R. 2014, Modern Physics Letters A, 29,
  1430002, publisher: World Scientific Publishing Co.

\bibitem[{Rogers \& Peiris(2021)}]{Rogers2021}
Rogers, K.~K. \& Peiris, H.~V. 2021, Physical Review Letters, 126, 071302,
  publisher: American Physical Society

\bibitem[{Roszkowski {et~al.}(2018)Roszkowski, Sessolo, \&
  Trojanowski}]{Roszkowski2018}
Roszkowski, L., Sessolo, E.~M., \& Trojanowski, S. 2018, Reports on Progress in
  Physics, 81, 066201, publisher: IOP Publishing

\bibitem[{Saikawa \& Yamaguchi(2013)}]{Saikawa2013}
Saikawa, K. \& Yamaguchi, M. 2013, Physical Review D, 87, 085010, publisher:
  American Physical Society

\bibitem[{Sakharov(1966)}]{Sakharov1966}
Sakharov, A.~D. 1966, Soviet Journal of Experimental and Theoretical Physics,
  22, 241, aDS Bibcode: 1966JETP...22..241S

\bibitem[{Schive {et~al.}(2014)Schive, Chiueh, \& Broadhurst}]{Schive2014}
Schive, H.-Y., Chiueh, T., \& Broadhurst, T. 2014, Nature Physics, 10, 496

\bibitem[{Scolnic {et~al.}(2018)Scolnic, Jones, Rest, Pan, Chornock, Foley,
  Huber, Kessler, Narayan, Riess, Rodney, Berger, Brout, Challis, Drout,
  Finkbeiner, Lunnan, Kirshner, Sanders, Schlafly, Smartt, Stubbs, Tonry,
  Wood-Vasey, Foley, Hand, Johnson, Burgett, Chambers, Draper, Hodapp, Kaiser,
  Kudritzki, Magnier, Metcalfe, Bresolin, Gall, Kotak, McCrum, \&
  Smith}]{Scolnic2018}
Scolnic, D.~M., Jones, D.~O., Rest, A., {et~al.} 2018, The Astrophysical
  Journal, 859, 101, publisher: American Astronomical Society

\bibitem[{Shapiro {et~al.}(2021)Shapiro, Dawoodbhoy, \&
  Rindler-Daller}]{Shapiro2021}
Shapiro, P.~R., Dawoodbhoy, T., \& Rindler-Daller, T. 2021, arXiv:2106.13244
  [astro-ph, physics:hep-ph], arXiv: 2106.13244

\bibitem[{Sikivie \& Yang(2009)}]{Sikivie2009}
Sikivie, P. \& Yang, Q. 2009, Physical Review Letters, 103, 111301

\bibitem[{Silverman \& Mallett(2002)}]{Silverman2002}
Silverman, M.~P. \& Mallett, R.~L. 2002, General Relativity and Gravitation,
  34, 633

\bibitem[{Steigman(2012)}]{Steigman2012}
Steigman, G. 2012, Advances in High Energy Physics, 2012, e268321, publisher:
  Hindawi

\bibitem[{Tegmark {et~al.}(2004)Tegmark, Blanton, Strauss, Hoyle, Schlegel,
  Scoccimarro, Vogeley, Weinberg, Zehavi, Berlind, Budavari, Connolly,
  Eisenstein, Finkbeiner, Frieman, Gunn, Hamilton, Hui, Jain, Johnston, Kent,
  Lin, Nakajima, Nichol, Ostriker, Pope, Scranton, Seljak, Sheth, Stebbins,
  Szalay, Szapudi, Verde, Xu, Annis, Bahcall, Brinkmann, Burles, Castander,
  Csabai, Loveday, Doi, Fukugita, III, Hennessy, Hogg, Ivezi{\'{c}}, Knapp,
  Lamb, Lee, Lupton, McKay, Kunszt, Munn, O'Connell, Peoples, Pier, Richmond,
  Rockosi, Schneider, Stoughton, Tucker, Berk, Yanny, \& and}]{Tegmark2004}
Tegmark, M., Blanton, M.~R., Strauss, M.~A., {et~al.} 2004, The Astrophysical
  Journal, 606, 702

\bibitem[{Thomas {et~al.}(2019)Thomas, Kopp, \& Markovič}]{Thomas2019}
Thomas, D.~B., Kopp, M., \& Markovič, K. 2019, Monthly Notices of the Royal
  Astronomical Society, 490, 813

\bibitem[{Thomas {et~al.}(2016)Thomas, Kopp, \& Skordis}]{Thomas2016}
Thomas, D.~B., Kopp, M., \& Skordis, C. 2016, 830, 155, publisher: American
  Astronomical Society

\bibitem[{Trujillo-Gomez {et~al.}(2011)Trujillo-Gomez, Klypin, Primack, \&
  Romanowsky}]{Trujillo-Gomez2011}
Trujillo-Gomez, S., Klypin, A., Primack, J., \& Romanowsky, A.~J. 2011, The
  Astrophysical Journal, 742, 16, publisher: American Astronomical Society

\bibitem[{Velten \& Wamba(2012)}]{Velteen2012}
Velten, H. \& Wamba, E. 2012, Physics Letters B, 709, 1

\bibitem[{Villanueva-Domingo {et~al.}(2021)Villanueva-Domingo, Mena, \&
  Palomares-Ruiz}]{Villanueva-Domingo2021}
Villanueva-Domingo, P., Mena, O., \& Palomares-Ruiz, S. 2021, Frontiers in
  Astronomy and Space Sciences, 8, publisher: Frontiers

\bibitem[{Vogelsberger {et~al.}(2014)Vogelsberger, Genel, Springel, Torrey,
  Sijacki, Xu, Snyder, Bird, Nelson, \& Hernquist}]{Vogelsberger2014}
Vogelsberger, M., Genel, S., Springel, V., {et~al.} 2014, Nature, 509, 177,
  number: 7499 Publisher: Nature Publishing Group

\bibitem[{Weinberg {et~al.}(2015)Weinberg, Bullock, Governato, Naray, \&
  Peter}]{Weinberg2015}
Weinberg, D.~H., Bullock, J.~S., Governato, F., Naray, R. K.~d., \& Peter, A.
  H.~G. 2015, Proceedings of the National Academy of Sciences, 112, 12249,
  publisher: National Academy of Sciences Section: Colloquium Paper

\bibitem[{Zhang {et~al.}(2018)Zhang, Chan, Harko, Liang, \& Leung}]{Zhang2018}
Zhang, X., Chan, M.~H., Harko, T., Liang, S.-D., \& Leung, C.~S. 2018, The
  European Physical Journal C, 78, 346

\end{thebibliography}

\end{document}